\documentclass[aps,twocolumn,groupedaddress,superscriptaddress]{revtex4-1}
\usepackage{graphicx,amsmath,amssymb}

\begin{document}

\begin{titlepage}
\title{The yielding transition in amorphous solids under oscillatory shear deformation}
\author{Premkumar Leishangthem} 
\affiliation{Jawaharlal Nehru Center for Advanced Scientific Research, Jakkur Campus, Bengaluru 560064, India}
\author{Anshul D. S. Parmar}
\affiliation{Jawaharlal Nehru Center for Advanced Scientific Research, Jakkur Campus, Bengaluru 560064, India}
\affiliation{TIFR Center for Interdisciplinary Sciences, 21 Brundavan Colony, Narsingi, Hyderabad  500075, India}
\author{Srikanth Sastry}
\affiliation{Jawaharlal Nehru Center for Advanced Scientific Research, Jakkur Campus, Bengaluru 560064, India}


\begin{abstract}
  Amorphous solids are ubiquitous among natural and man-made
  materials. Often used as structural materials for their attractive
  mechanical properties, their utility depends critically on their
  response to applied stresses. Processes underlying such mechanical
  response, and in particular the yielding behaviour of amorphous
  solids, are not satisfactorily understood. Although studied
  extensively\cite{Falk2010,Barrat2011,Maloney2006,Dahmen2009,Karmakar2010,Dasgupta2012,Keim2013,Lin2014,Knowlton2014,Nagamanasa2014,Shrivastav2016,regev2015reversibility,Liu2016,Itamar2016yielding},
  observed yielding behaviour can be gradual and depend significantly on conditions
  of study, making it difficult to convincingly validate existing
  theoretical descriptions of a sharp yielding transition\cite{Hebraud1998,Dahmen2009,Dasgupta2012,Lin2014,Liu2016}. Here, we
  employ oscillatory deformation as a reliable probe of
  the yielding transition. Through extensive computer simulations for
  a wide range of system sizes, we demonstrate that cyclically
  deformed model glasses exhibit a sharply defined yielding transition
  with characteristics that are independent of preparation history. In
  contrast to prevailing
  expectations\cite{Dahmen2009,Lin2014,regev2015reversibility}, the
  statistics of avalanches reveals no signature of the impending
  transition, but exhibit dramatic, qualitative, changes in character
  across the transition.
\end{abstract}

\maketitle
\end{titlepage}


The mechanical response to applied stresses or deformation is a basic
material characteristic of solids, both crystalline and
amorphous. Whereas the response to small perturbations are described
by elastic moduli, the {\it plastic}, irreversible, response to large
deformation is often more important to characterise, as it determines
many material parameters such as strength and ductility, and is also
of relevance to thermomechanical processing of metallic glasses \cite{Greer2016}. 
Amorphous solids lack the translational symmetry of crystals, and thus no obvious analogs to
dislocation defects in terms of which plasticity in crystals has been sought to be understood. 
Based on work over the last decades, it is appreciated that
plasticity arises in amorphous solids through spatially localized
reorganisations \cite{Argon1979,Falk2010,Barrat2011}, termed shear
transformation zones, and that such localized zones interact with each
other through long ranged elastic strains they
induce \cite{Picard2004}. While many details of the nature of these
localized regions of {\it non-affine} displacements remain to be
worked out, they form the basis of analyses and models of {\it
  elasto-plasticity} and yielding \cite{Dasgupta2012,Talamali2012,Picard2004,Liu2016,Surajit2016}. 
Many analyses have employed computer simulations
of atomistic models of glasses, aiming to elucidate key features of plastic response
\cite{Maloney2006,Falk2010,Barrat2011} on atomic scales. 
While several
studies have been conducted at finite shear rates ({\it e.g.}
\cite{Shrivastav2016,Liu2016}), many studies have focussed on
behaviour in the athermal, quasi-static (AQS)
\cite{Maloney2006,Karmakar2010,Dasgupta2012,Itamar2016yielding,Salerno2013}
limit, wherein the model glasses studied remain in zero temperature,
local energy minimum, configurations as they are sheared
quasi-statically. Such deformation induces discontinuous drops in energy and stress 
with corresponding nonaffine displacements that are highly spatially correlated, and exhibit power law distributions in size. 
In analogy with similar {\it avalanches} that arise in diverse context of intermittent response in disordered
systems, from earthquakes, crackling noise in magnetic systems,
depinning of interfaces in a disorded medium {\it etc.} \cite{Sethna2001}, a theoretical description of yielding in amorphous
solids \cite{Dahmen2009}, predicts the mean avalanche size to diverge as the
yielding transition is approached from below, leading to a power law
distribution with a diverging mean size at and above the transition. 
Indeed, it has been observed that ({\it e. g.}
\cite{Karmakar2010,Salerno2013,Liu2016}) system spanning avalanches
are present in the steady state beyond yield, whose sizes scale with
system size. The character of avalanches upon approaching the yielding
transition, however, has not received much attention, as also the
differences between pre- and post-yield avalanches. Among the reasons is the sample to sample variability of behaviour below
yield, in contrast with the universal behaviour seen in the post-yield
regime. Here, we show that oscillatory deformation offers a robust
approach to systematically probe behaviour above and below
yielding. Oscillatory deformation is a widely used experimental
technique as well as a common protocol in materials testing. However, barring some recent work \cite{Fiocco2013,Priezjev2013,Regev2013,regev2015reversibility}, 
it is not been employed widely to probe yielding
 in amorphous solids computationally.  In
the present work, we perform an extensive computational study of
plastic response in a model glass former, over a wide range of system
sizes, and amplitudes of deformation that straddle the yielding
strain.

We study the Kob-Andersen 80:20 binary mixture Lennard-Jones glasses for a range of system sizes 
(see Methods for details). The glasses studied are
prepared by performing a local energy minimization of equilibrated
liquid configurations, at a reduced temperatures $T = 1$ and $T =
0.466$. The {\it inherent
  structures} so obtained represent poorly annealed ($T = 1$) and well
annealed ($T = 0.466$) glasses. 
 These glasses, referred to by the  corresponding liquid temperature in what follows, 
are subjected to volume preserving shear deformation through
the AQS protocol. The strain is
incremented in the same direction in the case of {\it uniform strain}, whereas for oscillatory strain for
a given maximum amplitude $\gamma_{max}$, a cycle of strain $0
\rightarrow \gamma_{max} \rightarrow 0 \rightarrow -\gamma_{max}
\rightarrow 0$ is applied repeatedly over many cycles, till a steady
state is reached. Results presented below, except Figure 1 (d) are from analysing steady state configurations. 
Further details concerning the simulations and
analysis are presented in Methods and Supporting Information.

{\it Stress and Energy across the Yielding Transition.-}
Previous work \cite{Fiocco2013} has shown that as the
amplitude of strain $\gamma_{max}$ approaches a critical value
$\gamma_y$ from either side, the approach to the steady state becomes
increasingly sluggish, with an apparent divergence at $\gamma_y$ (see 
Supporting Information).  We identify $\gamma_y$
($\sim 0.07$) as the yield strain, as justified below. In
Figure 1(a) we show the averaged stress-strain curves for $N = 4000$.
For each $\gamma_{max}$, we obtain a maximum stress $\sigma_{max}$
reached at $\gamma = \gamma_{max}$, which are plotted in Figure 1(b)
for $T = 1$, $0.466$, for $N = 4000, 32000$. Figure 1(b) also shows
the stress-strain curves for the same cases obtained with uniform
strain. Whereas stresses vary smoothly for uniform strain, 
with no sharp signature of the onset of yielding, and 
differ significantly for $T = 1$ and $T = 0.466$, they display 
 a sharp, discontinuous, drop above $\gamma_{max} = 0.07$ ($0.08$
for N = 4000)  for oscillatory strain. Interestingly, below $\gamma_y$, the maximum stress
increases as a result of oscillatory deformation, indicative of
hardening, consistently with previous results \cite{Schuh2012}. Above
$\gamma_y$, repeated
oscillatory deformation leads to a stress drop relative to values just below $\gamma_y$, indicating yielding. Figure
1 (c) displays the potential energies obtained over a full cycle in
the steady state (see Supporting Information for evolution with cycles of strain). For $\gamma_{max} < \gamma_y$, the energies display
a single minimum close to $\gamma = 0$, but above, bifurcate into two
minima, indicating the emergence of plasticity. The stress-strain
curves show a corresponding emergence of loops (Figure 1(a)) with finite
area. Strain values at the minima
for energy, $\gamma_{Umin}$ and $\sigma_{xz} = 0$, $\gamma_{\sigma_0}$,
are shown in Figure 1(d) as a function of the number of cycles for
different $\gamma_{max}$. We note that $\gamma_{max} = 0.08$ displays
interesting non-monotonic behaviour, with an initial decrease in these
strain values, similar to smaller $\gamma_{max}$, but an eventual
increase to larger strains, similar to the case $\gamma_{max}= 0.12$,
in the yielded regime. Figure 1(e) shows $\gamma_{Umin}$ and
$\gamma_{\sigma_0}$ {\it vs.} $\gamma_{max}$, which show an apparently
continuous departure from nearly zero, signalling a transition at
$\gamma_{max} \sim 0.07$. Figure 1(f) shows that the minimum energies
in the steady state {\it vs.} $\gamma_{max}$ decrease with increasing
$\gamma_{max}$ below $\gamma_y$, but increase above, reaching the same
values for $T = 1$ and $T = 0.466$. These data demonstrate the
presence of a sharp transition between a low strain regime where
oscillatory shear produces better annealed, hardened, glasses to a
yielded regime displaying stress relaxation and rejuvenation.

\begin{figure*}[htp]
\centering  
\includegraphics[width=0.95\textwidth]{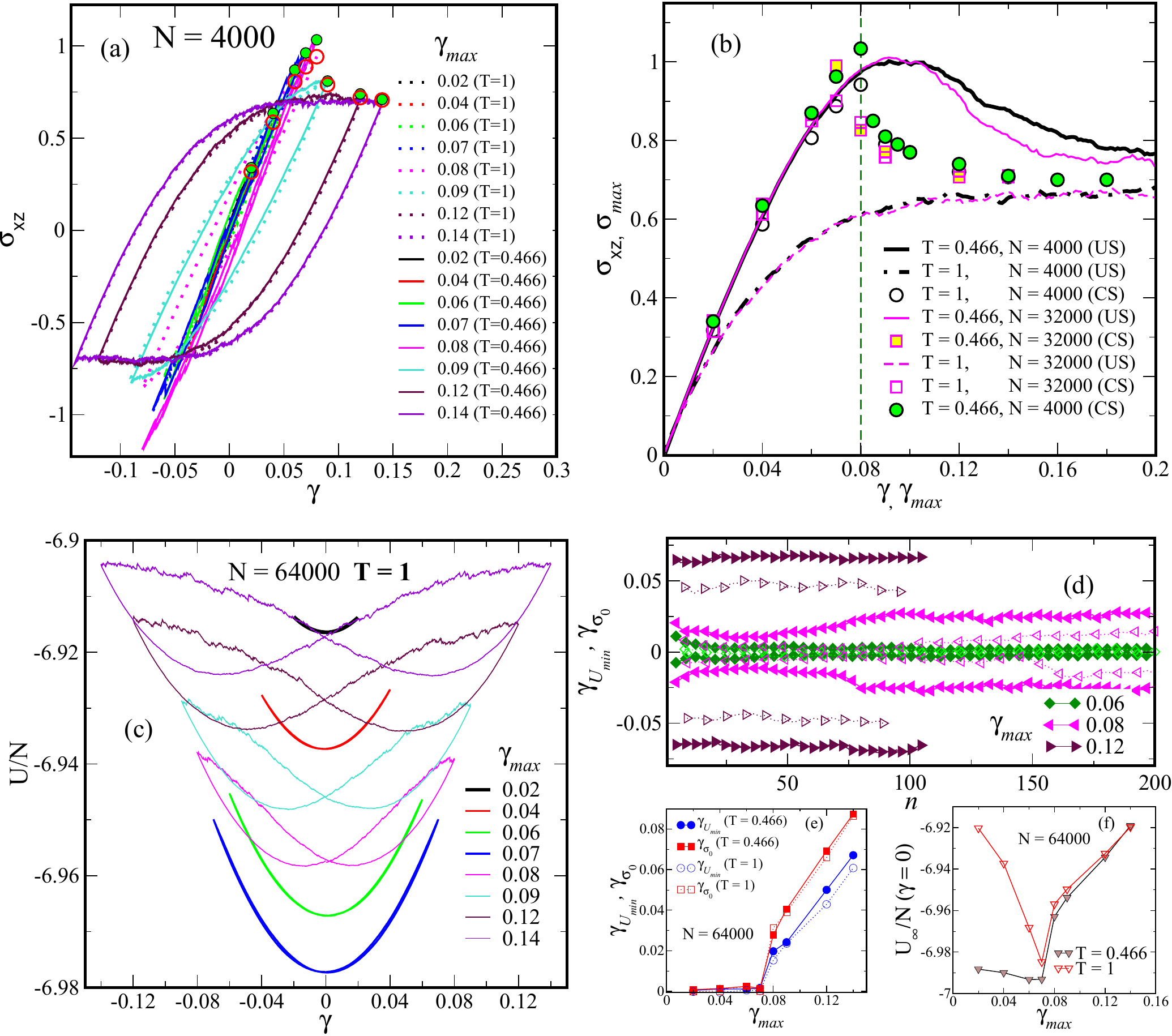}
\caption{Stress and energy across the yielding transition: {\bf (a)} Stress-strain plots of the two
differently annealed glasses for various strain amplitudes in the
steady states of oscillatory shear deformation.  Maximum stress in the
cycle for each amplitude is marked by filled and open circles for T =
0.466 and T = 1 respectively.  {\bf (b)} Averaged stress-strain curves
for uniform strain (US) are shown as lines - thick (black) for N=4000
and thin (magenta) for N = 32000 while solid and dashed lines
represent T=0.466 and T=1 respectively. Maximum stress $\sigma_{max}$
{\it vs.} $\gamma_{max}$ are shown for cyclic strain (CS) (circle and
square denote N = 4000 and 32000 respectively, with filled and open
symbols corresponding to glasses from T = 0.466 and T = 1).  The
vertical line at $\gamma_{max}=0.08$ indicates the sharp yielding
transition seen.  {\bf (c)} Energy {\it vs.} strain in the steady
states, displaying a bifurcation in the strain corresponding to minima
in energy at the yielding transition between $\gamma_{max} = 0.07$ and
$0.08$. {\bf (d)} Strain values corresponding to energy minima (
$\gamma_{U_{min}}$) and and zero stress ($\gamma_{\sigma_0}$) are
shown as open and filled symbols respectively, {\it vs.} the number of
cycles for different $\gamma_{max}$. For $\gamma_{max} = 0.08$ an
initial relaxation towards zero is reversed as the system evolves to
a yielded steady state with finite $\gamma_{U_{min}}$ and
$\gamma_{\sigma_0}$. {\bf (e)} $\gamma_{U_{min}}$ and
$\gamma_{\sigma_0}$ as functions of strain amplitude $\gamma_{max}$,
displaying a transition beyond $\gamma_{max}=0.07$.  {\bf (f)}
Asymptotic energy per particle at $\gamma = 0$ {\it vs.}  strain
amplitude $\gamma_{max}$. Energies decrease with $\gamma_{max}$ till
the yield strain is reached, after which they increase with
$\gamma_{max}$.}
\label{Fig1}
\end{figure*}

\begin{figure*}[htp]
\centering  
\includegraphics[width=.32\textwidth]{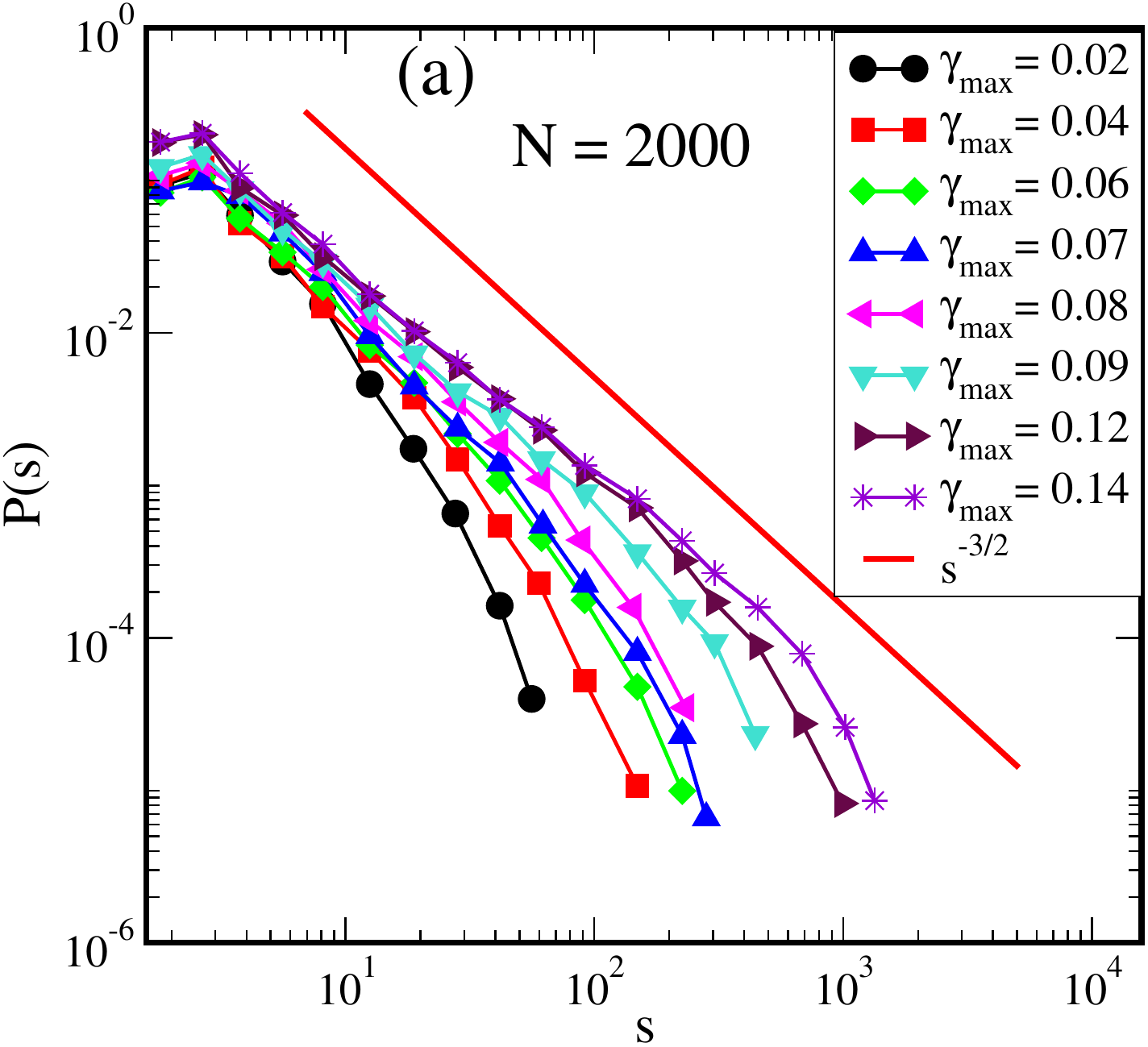}  
\includegraphics[width=.32\textwidth]{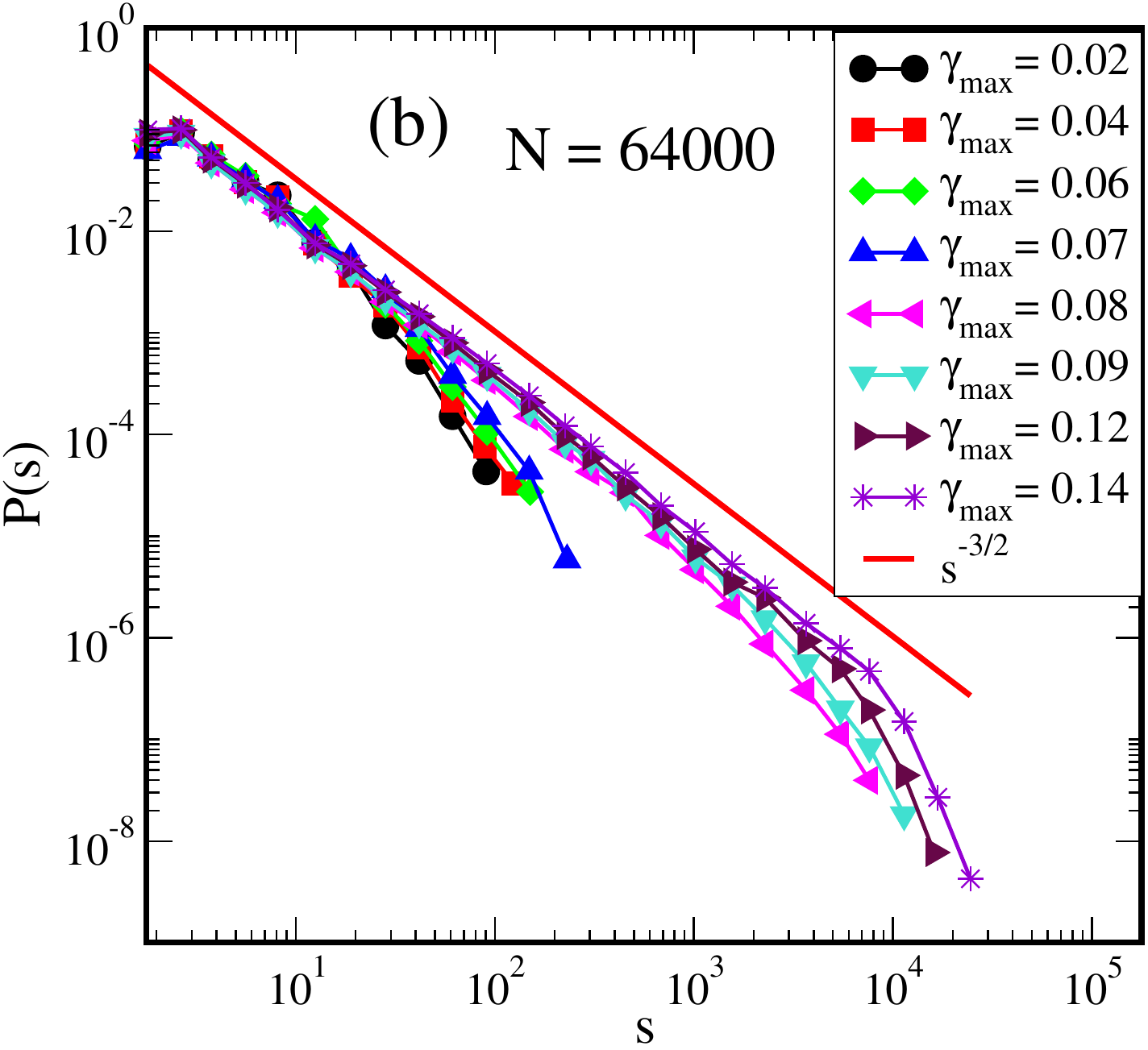} 
\includegraphics[width=.32\textwidth]{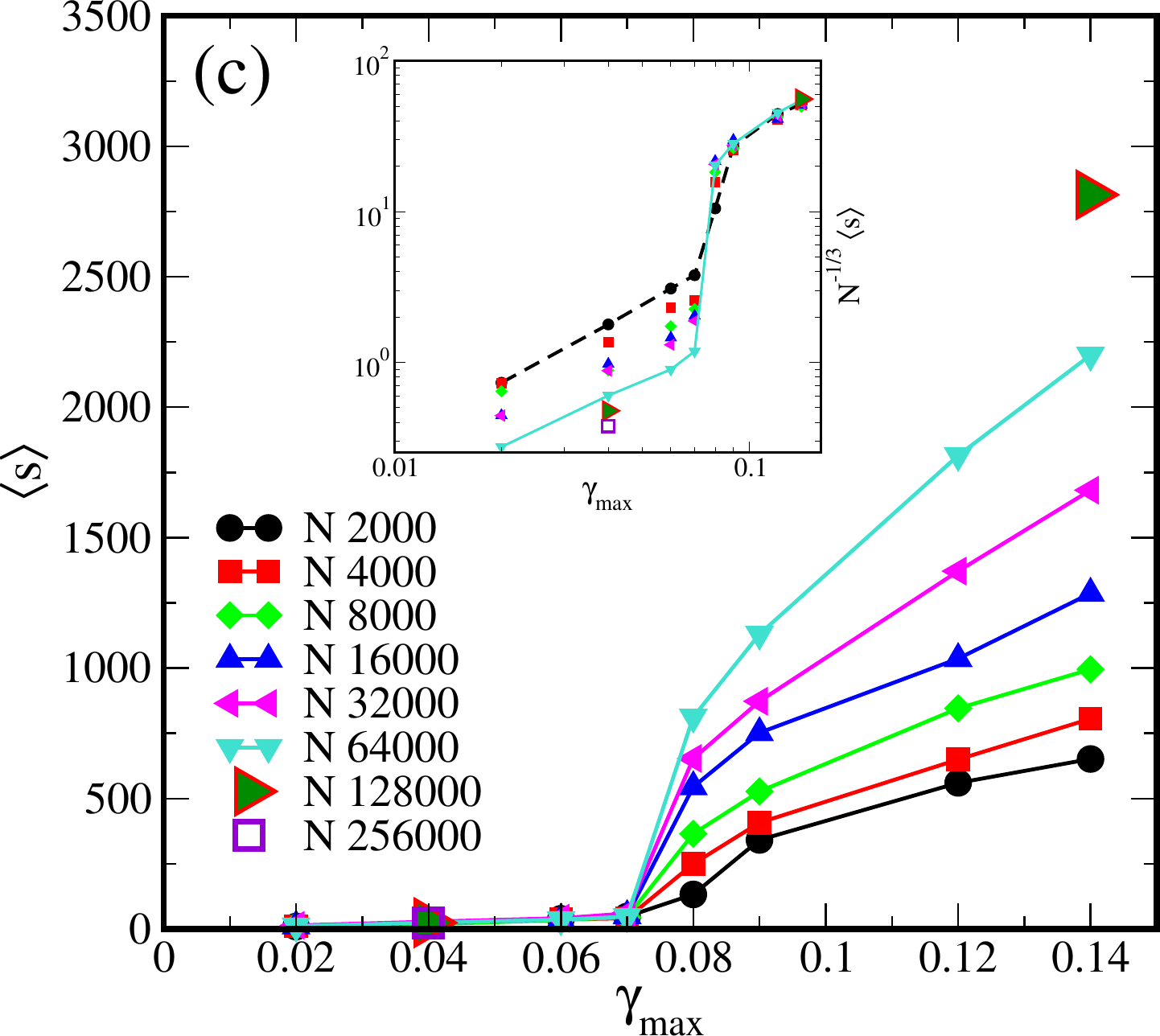} \\
\includegraphics[width=.32\textwidth]{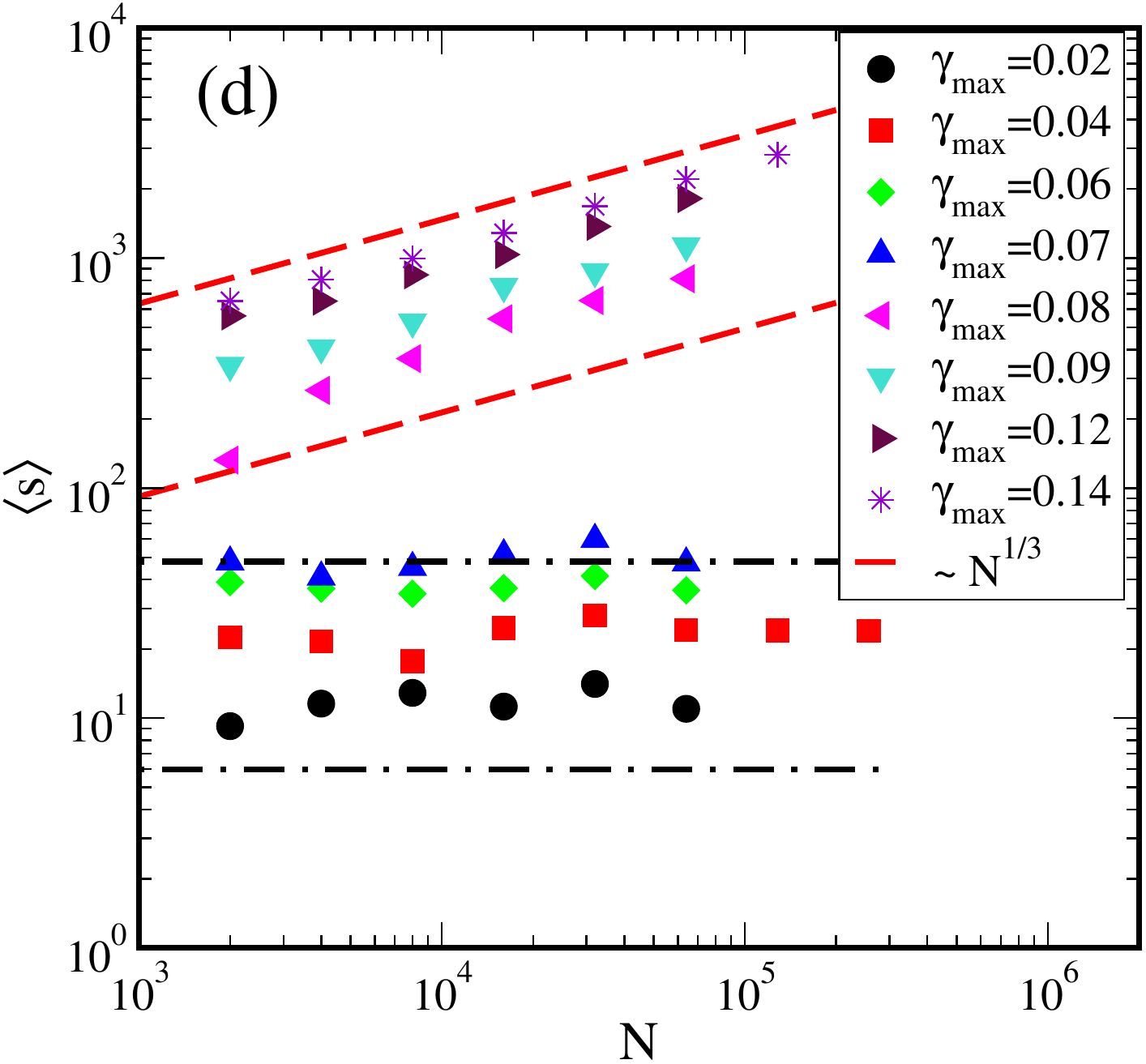}
\includegraphics[width=.325\textwidth]{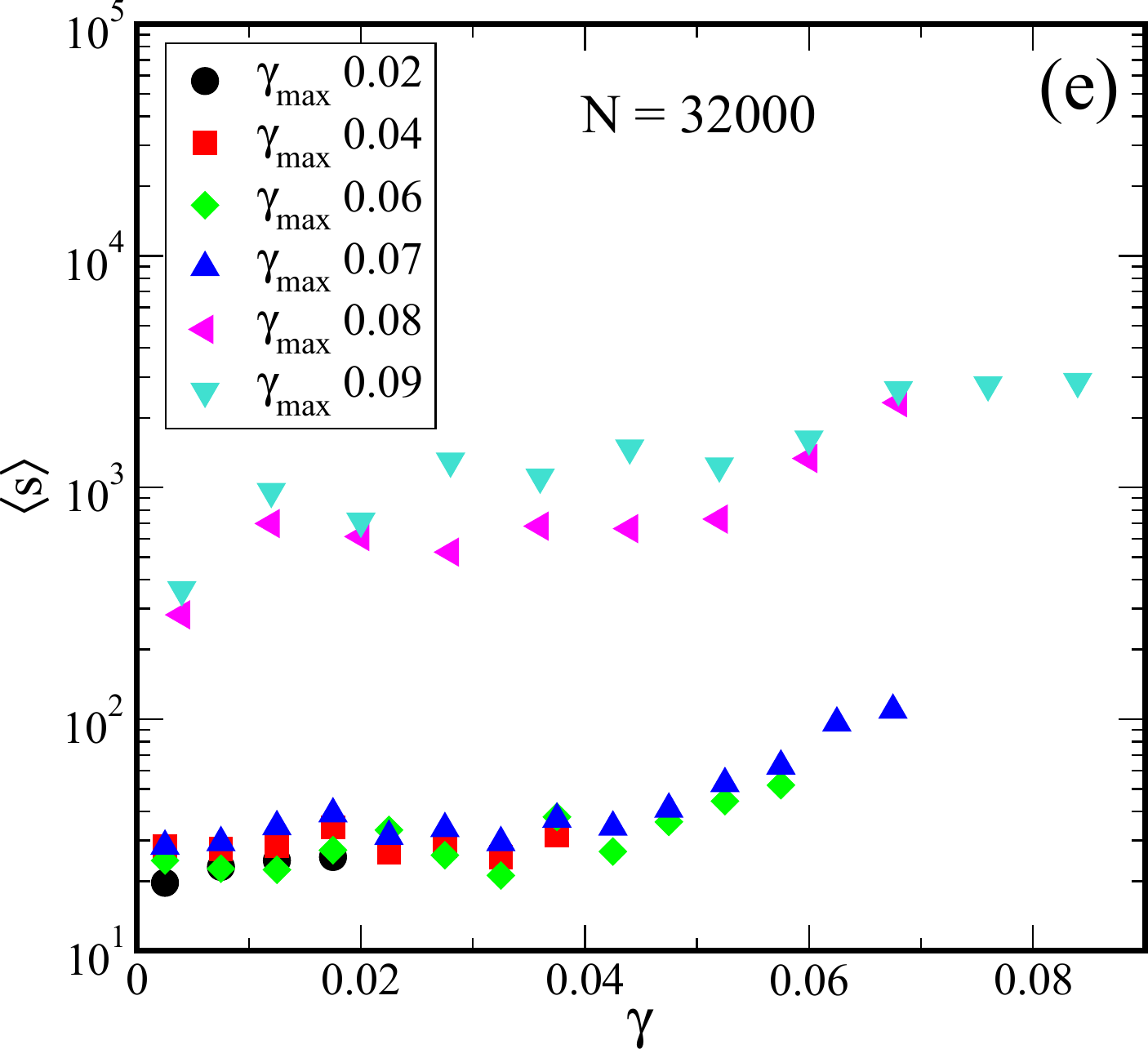}
\includegraphics[width=.33\textwidth]{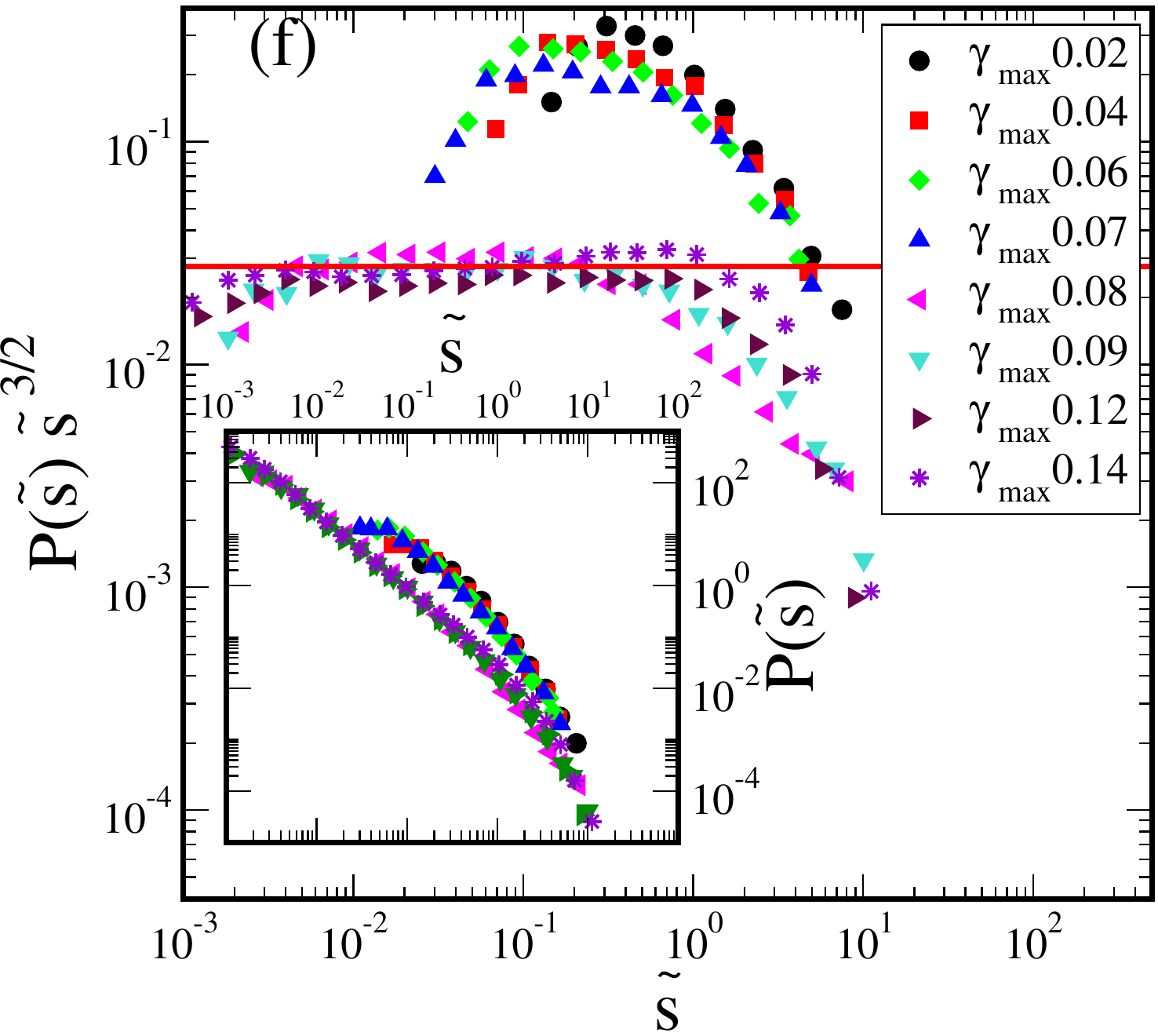}
\caption{ Statistics of avalanches as a function of
  strain amplitude $\gamma_{max}$ and system size $N$:  
  {\bf (a)}
Cluster size distributions for $N = 2000$ displaying a power law with a
cutoff that grows with $\gamma_{max}$ but does not indicate sharp
changes at yielding.  {\bf (b)} Cluster size distribution for $N =
64000$ displaying a sharp increase in the cutoff size across the
yielding transition. The line in both panels corresponds to a power
law with exponent $-3/2$.  {\bf (c)} Mean cluster size {\it vs.}
$\gamma_{max}$ showing a qualitative change across the yielding
transition, with strong system size dependence above $\gamma_y$. The
inset shows the mean cluster size scaled with $N^{1/3}$, which
describes well the size dependence above $\gamma_y$. {\bf (d)} Mean
cluster size {\it vs.} system size $N$ shows no significant size
dependence for $\gamma_{max} < \gamma_y$ but a clear $N^{1/3}$
dependence above. A crossover in behaviour is seen for $\gamma_{max} =
0.08$. Lines, with $N^{0}$ (constant) and $N^{1/3}$ dependence, are
guides to the eye.  {\bf (e)} Mean cluster sizes for bins in strain
$\gamma$ for different $\gamma_{max}$ for $N = 32000$.  Mean cluster size does not
depend on $\gamma_{max}$, and depends only mildly on strain $\gamma$,
for two distinct sets, below and above yield strain $\gamma_y$. {\bf
  (f)} Scaled cluster size ($\tilde{s} = s/<s>$) distributions exhibit 
data collapse separately for $\gamma_{max}  <  \gamma_y$ and
$\gamma_{max}  >  \gamma_y$ (inset). Distributions for $\gamma_{max}
 <  \gamma_y$ do not display a power law regime, whereas $\gamma_{max}
 >  \gamma_y$ do, over about two decades in $\tilde{s}$, as
highlighted in a plot of $P(\tilde{s}) \tilde{s}^{~3/2}$ {\it vs.}
$\tilde{s}$.  Data shown are for $T = 1$, and averages are over the full cycle, except for (e) which are averaged over the first quadrant. \\
 }
\end{figure*}

We next study  (i) distribution of  avalanche sizes, which we compute as
the size of clusters of particles that undergo plastic rearrangements
(see Methods for how they are identified), and (ii) distributions of the
size of energy drops. 

{\it Distributions of Cluster Sizes.-}
In Figure 2 (a) we show the distributions $P(s)$
of avalanche sizes $s$ for $N = 2000$, which display a
characteristic power law decay with a cutoff. Although the
cutoffs move to larger values as $\gamma_{max}$ increases, we see no
indication of a transition. To assess the role of system sizes, we
compute the avalanche sizes for a variety of system sizes. Figure 2(b)
shows the avalanche size distribution for $N = 64000$. The
distributions fall into two clear sets, corresponding to
$\gamma_{max}$ above and below $\gamma_y$. We compute and display in
Figure 2 (c) the mean avalanche size $<s>$ as a function of
$\gamma_{max}$, for all studied system sizes.  The striking
observation is that below $\gamma_y$, $<s>$ displays no system size
dependence, and only a very mild dependence on $\gamma_{max}$, and no
indication of the approach to $\gamma_y$. Above
$\gamma_y$, a clear system size dependence is seen. Figure 2 (d) shows
the same data {\it vs.} system size, revealing a roughly $N^{1/3}$
(or $<s> \sim L$) dependence above $\gamma_y$, and minimal $N$
dependence below. The $N^{1/3}$ dependence is consistent with previous
results \cite{Lerner2009,Karmakar2010} for mean energy drops, but
the absence of system size dependence below, to our knowledge, has
not been demonstrated before. We next ask whether the mean
size of avalanches, for a given $\gamma_{max}$ depend on the strain
$\gamma$ at which they appear, and conversely, for a given $\gamma$
what the dependence on $\gamma_{max}$ is. As shown in Figure 2 (e)
($N = 32000$, $T = 1$), for a given $\gamma_{max}$ the 
  $\gamma$ dependence is weak and is the same for  $\gamma_{max}
< \gamma_y$ (and $\gamma_{max} > \gamma_y$), but the data fall into distinct
groups for $\gamma_{max} < \gamma_y$ and $\gamma_{max} >
\gamma_y$. The same pattern is seen for the full distributions (see
Supporting Information). 
For a given $\gamma_{max}$, the avalanche distributions can be
collapsed on to a master curve by scaling $s$ by $<s>$ (data not
shown). The distributions of scaled sizes $\tilde{s} \equiv s/<s>$,
averaged over system size are shown in the inset of Figure 2 (f). The
same data are shown, multiplied by $\tilde{s}^{~3/2}$ in the
main panel, and demonstrate that the character of the distributions
are different above and below yield: whereas above $\gamma_y$ one
finds a range of sizes over which the power law form $P(s) \sim
s^{-3/2}$ is clearly valid (and thus the cutoff arises purely because of system size), below $\gamma_y$ this is not the case, and the qualitative shape of
the distributions is different (with a cutoff function multiplying the power law) \cite{Dahmen2009,Lin2014,Fiocco2014,regev2015reversibility}.

\begin{figure*}[t]
\centering  
\includegraphics[width=.32\textwidth]{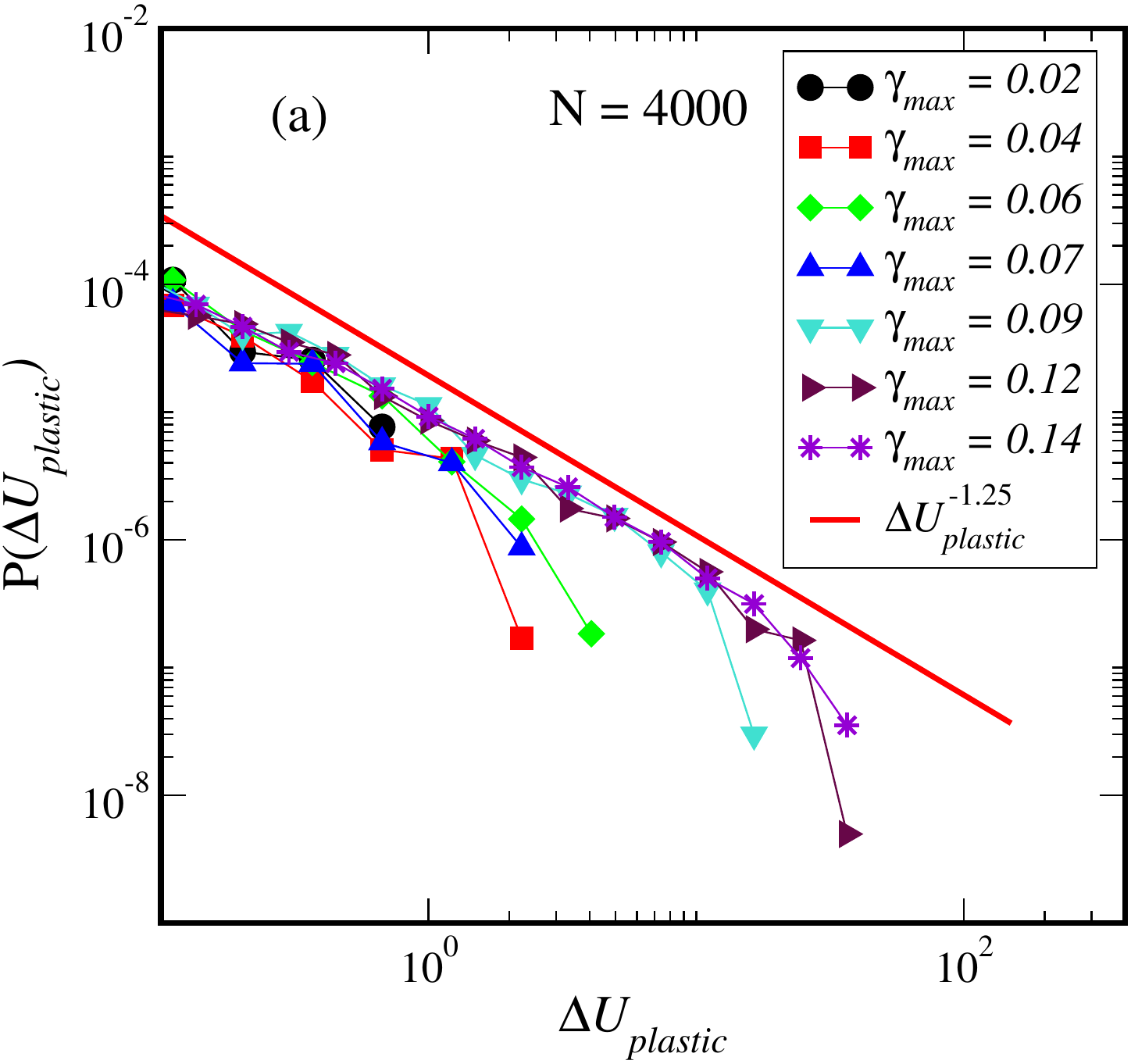} 
\includegraphics[width=.32\textwidth]{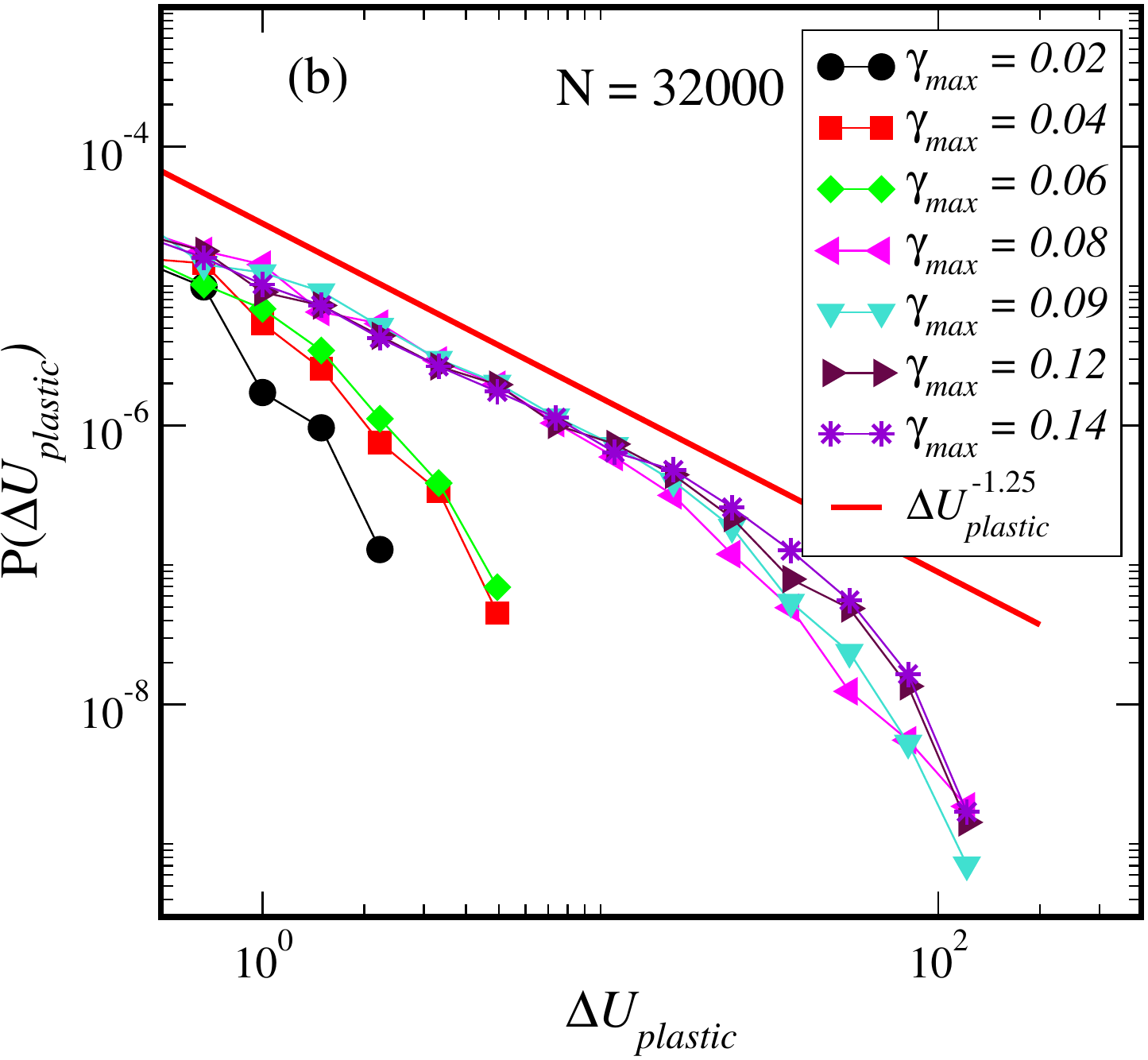}  
\includegraphics[width=.32\textwidth]{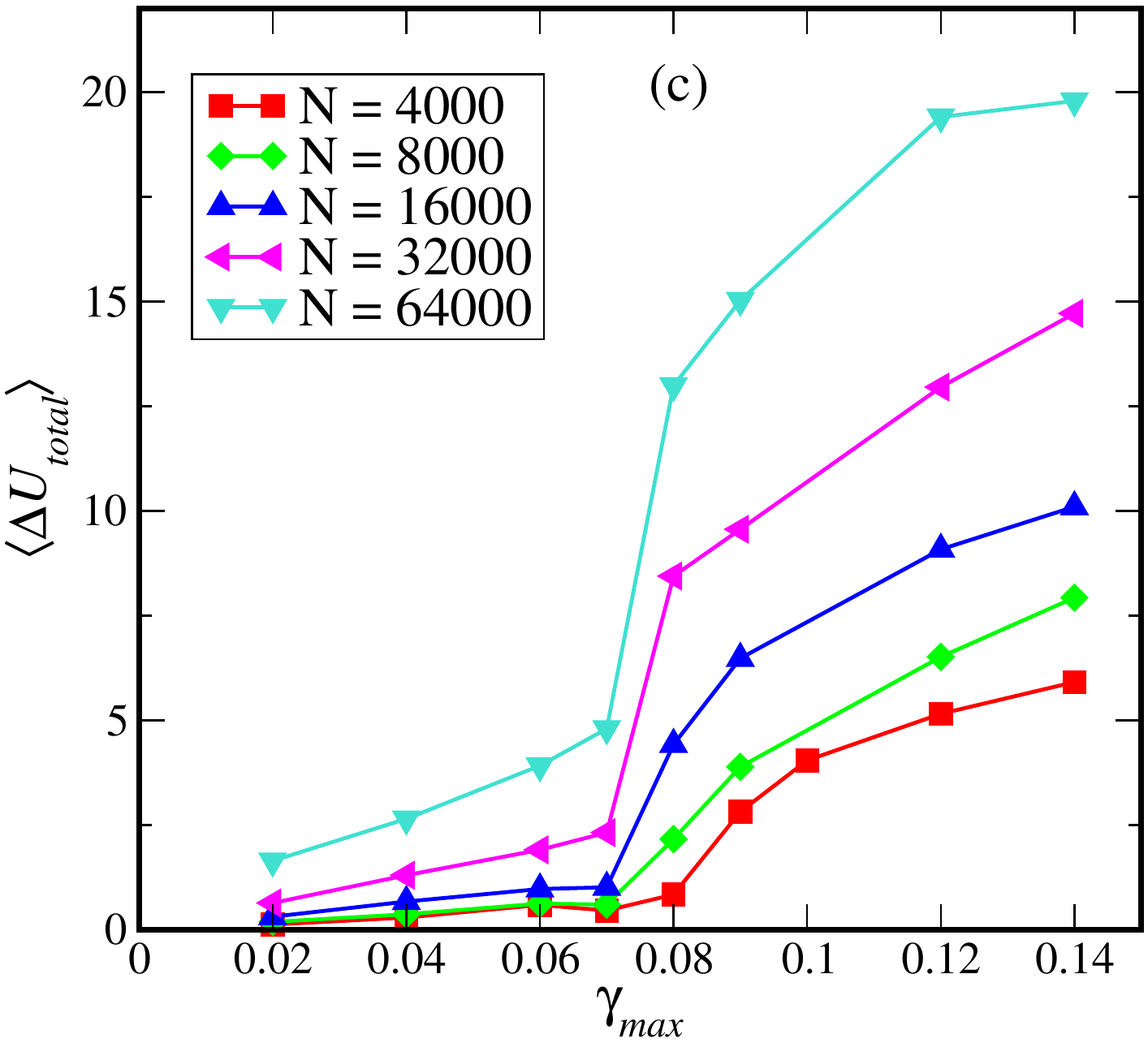} \\
\includegraphics[width=.32\textwidth]{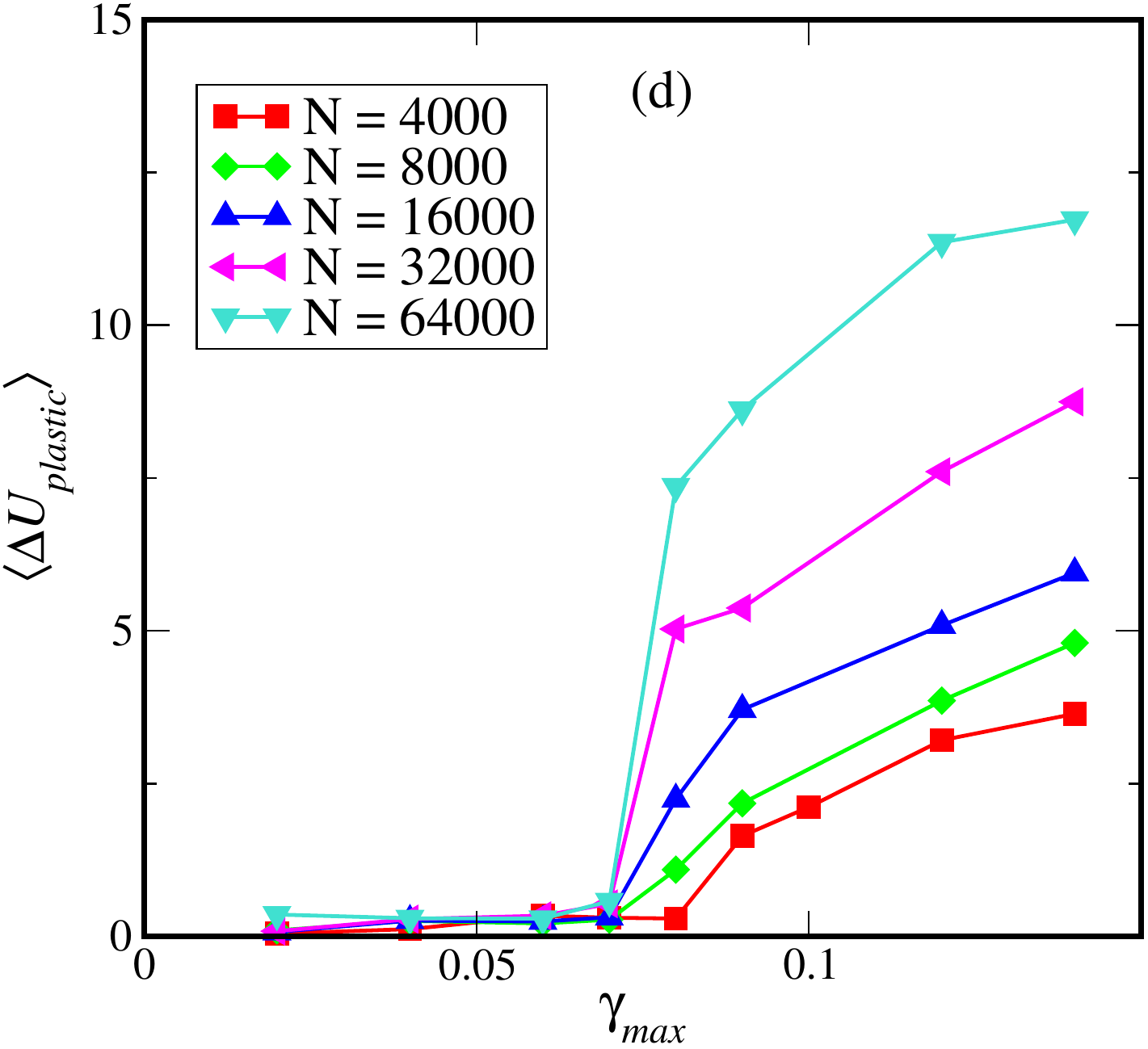}
\includegraphics[width=.32\textwidth]{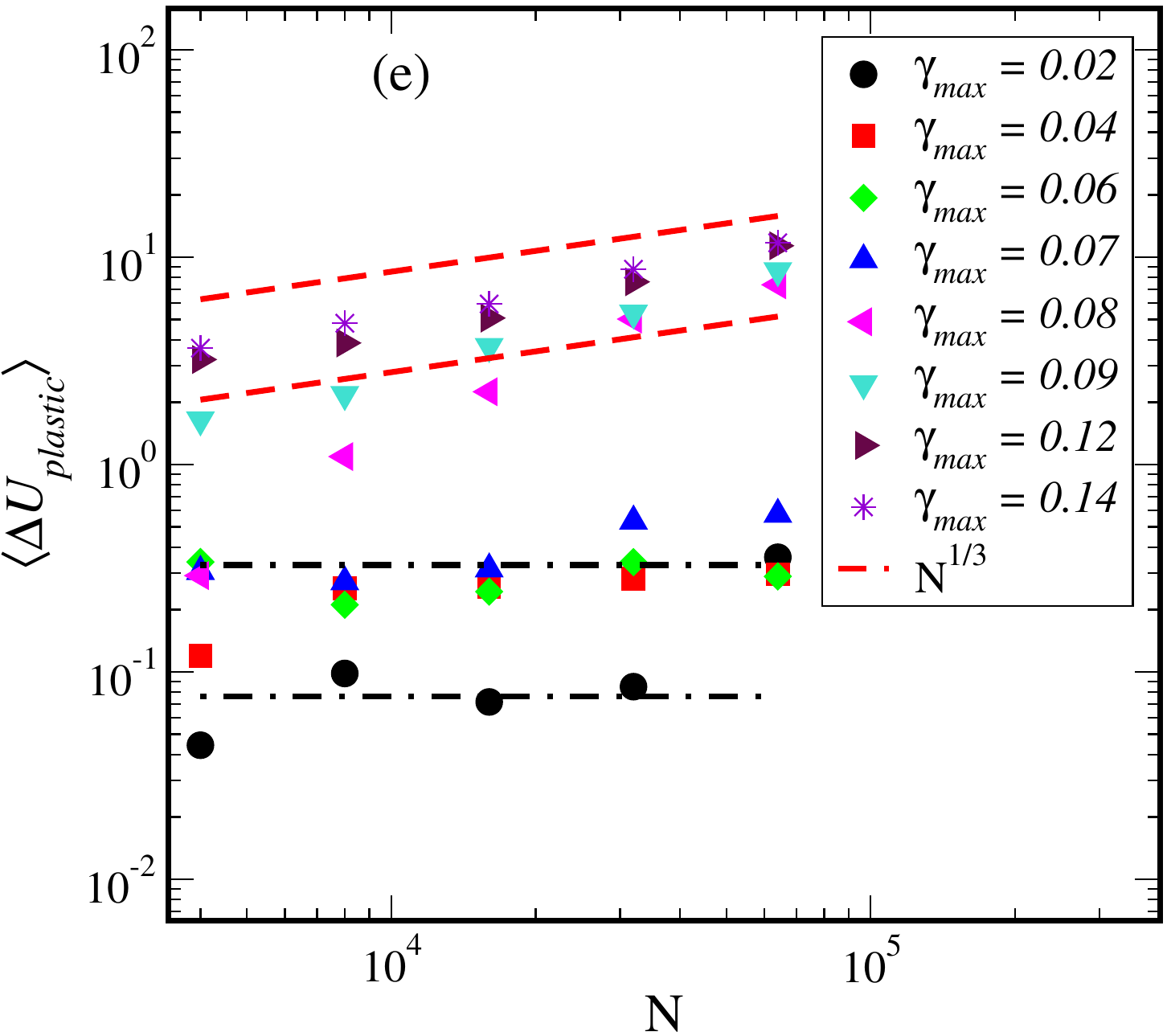} 
\includegraphics[width=.32\textwidth]{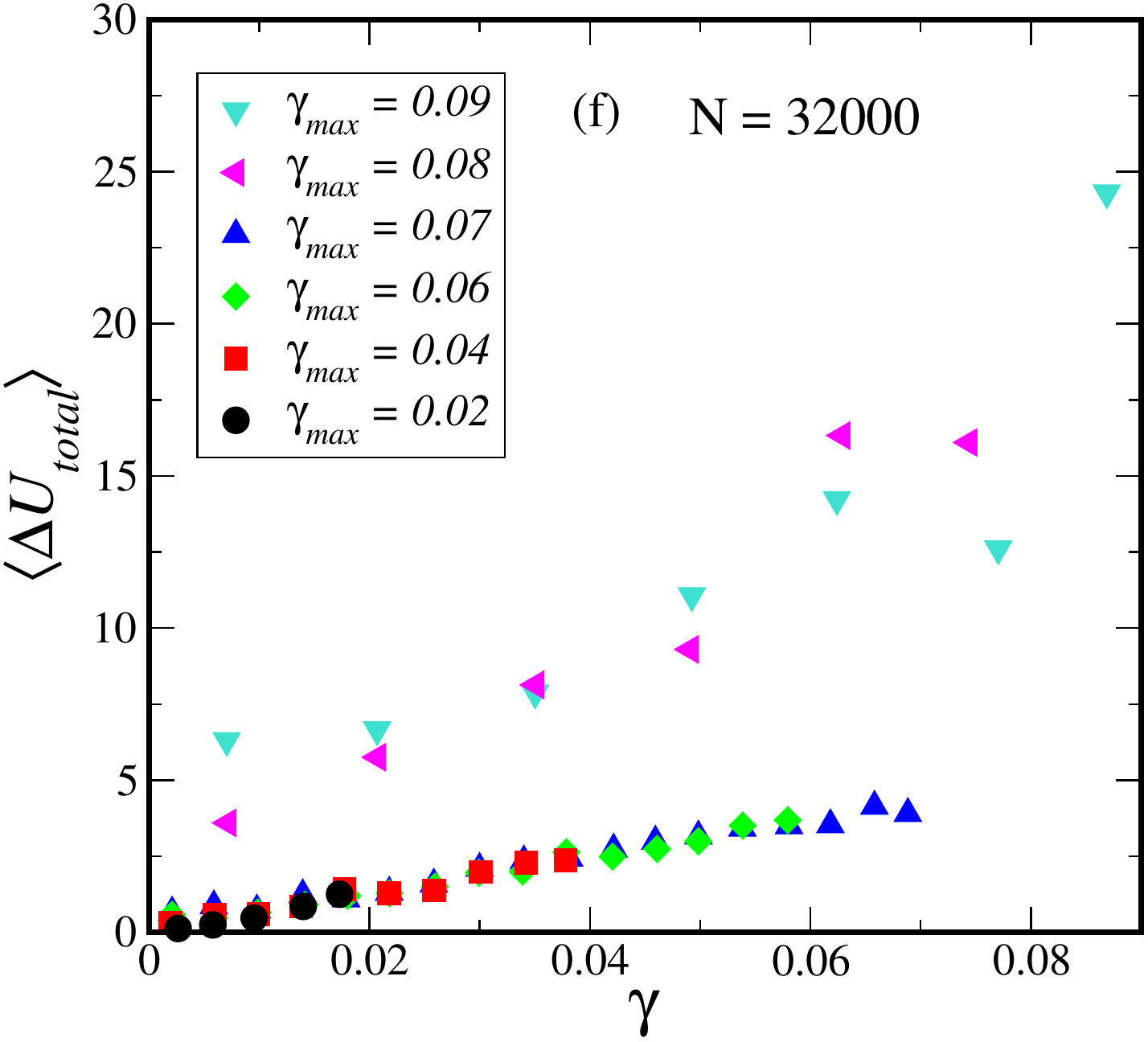}
\caption{Statistics of energy drops as a function of
  strain amplitude $\gamma_{max}$ and system size $N$: Distributions
of energy drops {\bf (a)} for $N = 4000$ show no clear separation of
$\gamma_{max} < \gamma_y$  and $\gamma_{max} > \gamma_y$  whereas {\bf (b)} for $N = 32000$ a
clear separation is visible. In both cases, a power law regime is
apparent, with exponent $\sim 1.25$.  {\bf (c)} Mean energy drops {\it
  vs.} $\gamma_{max}$, indicating a sharp change at $\gamma_y$. {\bf
  (d)} Mean energy drops considering only plastic regions show no
system size dependence below $\gamma_y$.  {\bf (e)} Mean energy drop
(plastic component) {\it vs.} system size $N$ shows no significant
size dependence for $\gamma_{max}  >  \gamma_y$ but a clear $N^{1/3}$
dependence above. A crossover in behaviour is seen for $\gamma_{max} =
0.08$. Lines, with $N^{0}$ (constant) and $N^{1/3}$ dependence, are
guides to the eye. {\bf (f)} Mean energy drops (total) for bins in
strain $\gamma$ for different $\gamma_{max}$ for $N = 32000$, $T = 1$
showing no dependence on $\gamma_{max}$, and only a mild dependence on
strain $\gamma$, for two distinct sets, below and above yield strain
$\gamma_y$. Data shown are for $T = 1$, and averages are  over the first quadrant. 
 }
\end{figure*}

{\it Distributions of Energy Drops.-}
We now discuss the distributions of energy drops. Shown for $N = 4000$
and $32000$ in Figures 3(a), and 3(b), these distributions show the
same features as the avalanche sizes, but with a different power law
exponent of $\sim 1.25$ (as found in \cite{Liu2016}. Thus, the
exponent depends on the quantity employed, and the avalanche size
based on particle displacements is in closer agreement with mean field
predictions). In Figure 3(c), we show the $\gamma_{max}$ dependence of
the mean energy drop, for different system sizes, which reveal the
same pattern as the avalanche sizes, albeit with a stronger apparent
size dependence below yield. However, the total energy drops for the whole system include also an
elastic component, in addition to the plastic component. The component of the energy
drop corresponding to the plastic regions alone, which are plotted in Figure
3 (d), to demonstrate that the plastic component has no system size
dependence below yield. Figure 3~(e) shows the system size dependence of the mean
energy drop (plastic component),  and Figure 3~(f) shows the mean energy drop {\it vs.}  $\gamma$ for different
$\gamma_{max}$ (N = 64000, T = 1), revealing the same separation below and above yield as the avalanche sizes. This is in contrast with the case of uniform shear, wherein both energy drops and avalanche sizes show a gradual, and strongly sample dependent, variation with strain (see Supporting Information). 

\begin{figure*}[t]
\centering 
\includegraphics[width=.26\textwidth]{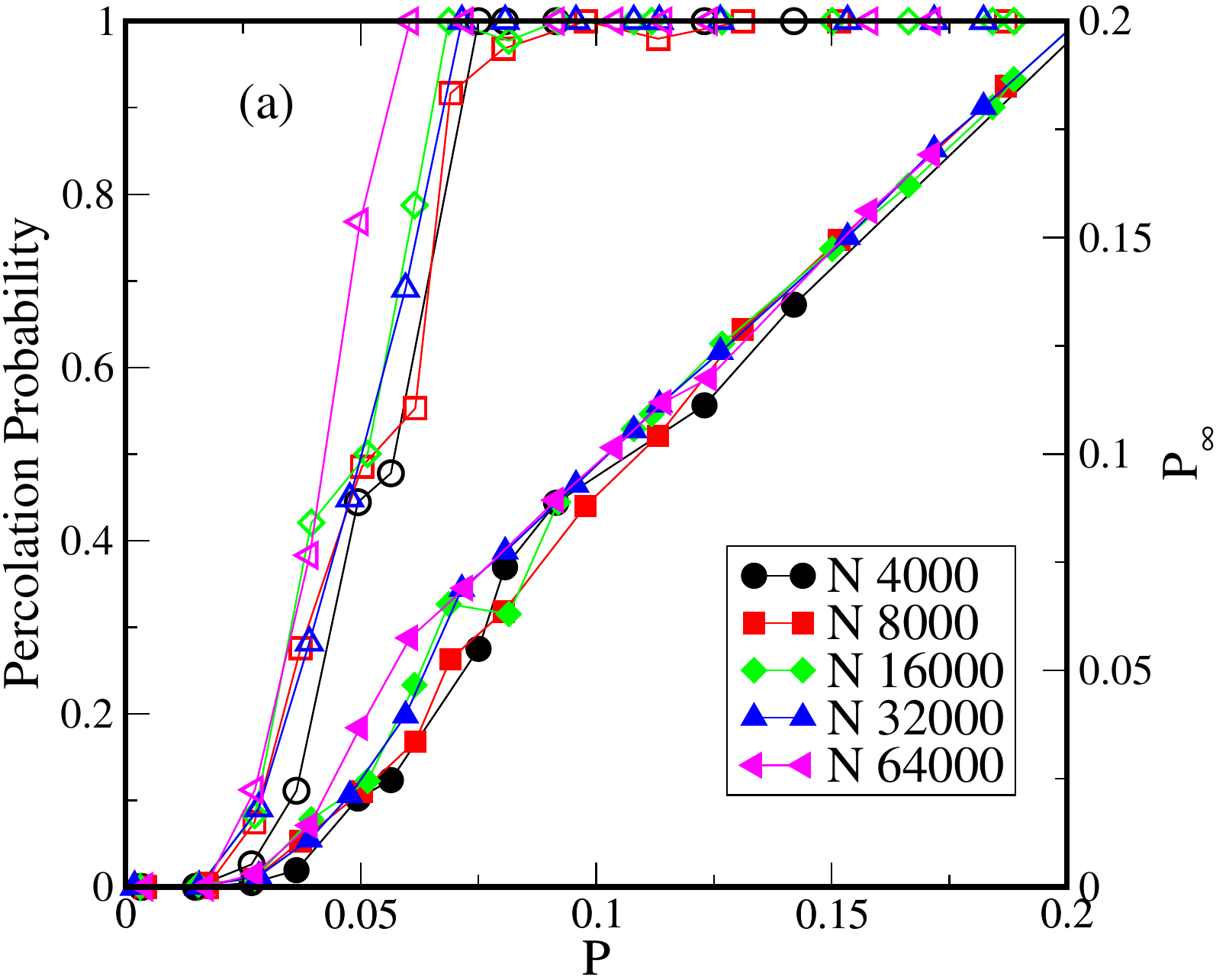} 
\includegraphics[width=.24\textwidth]{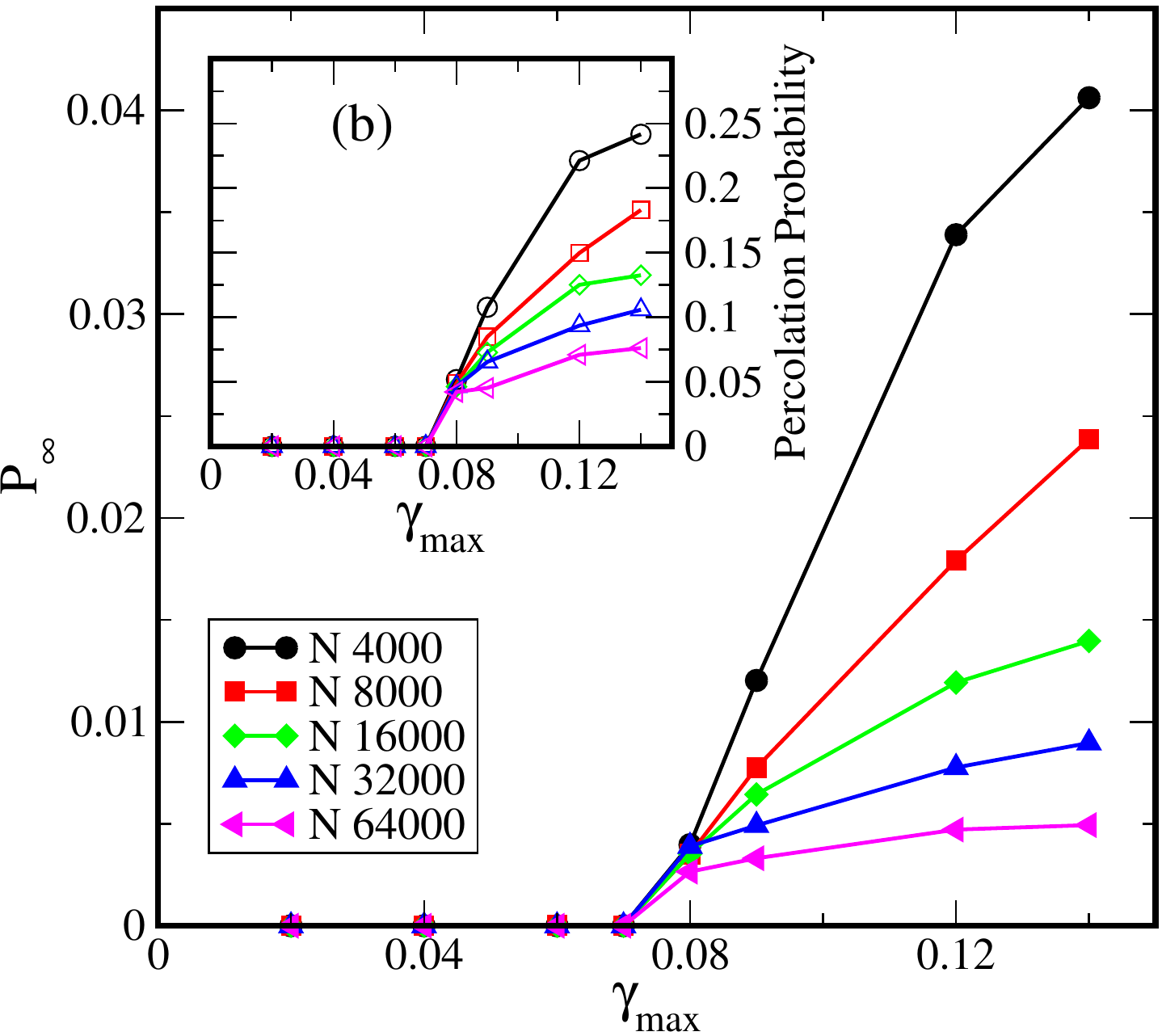}
\includegraphics[width=.24\textwidth]{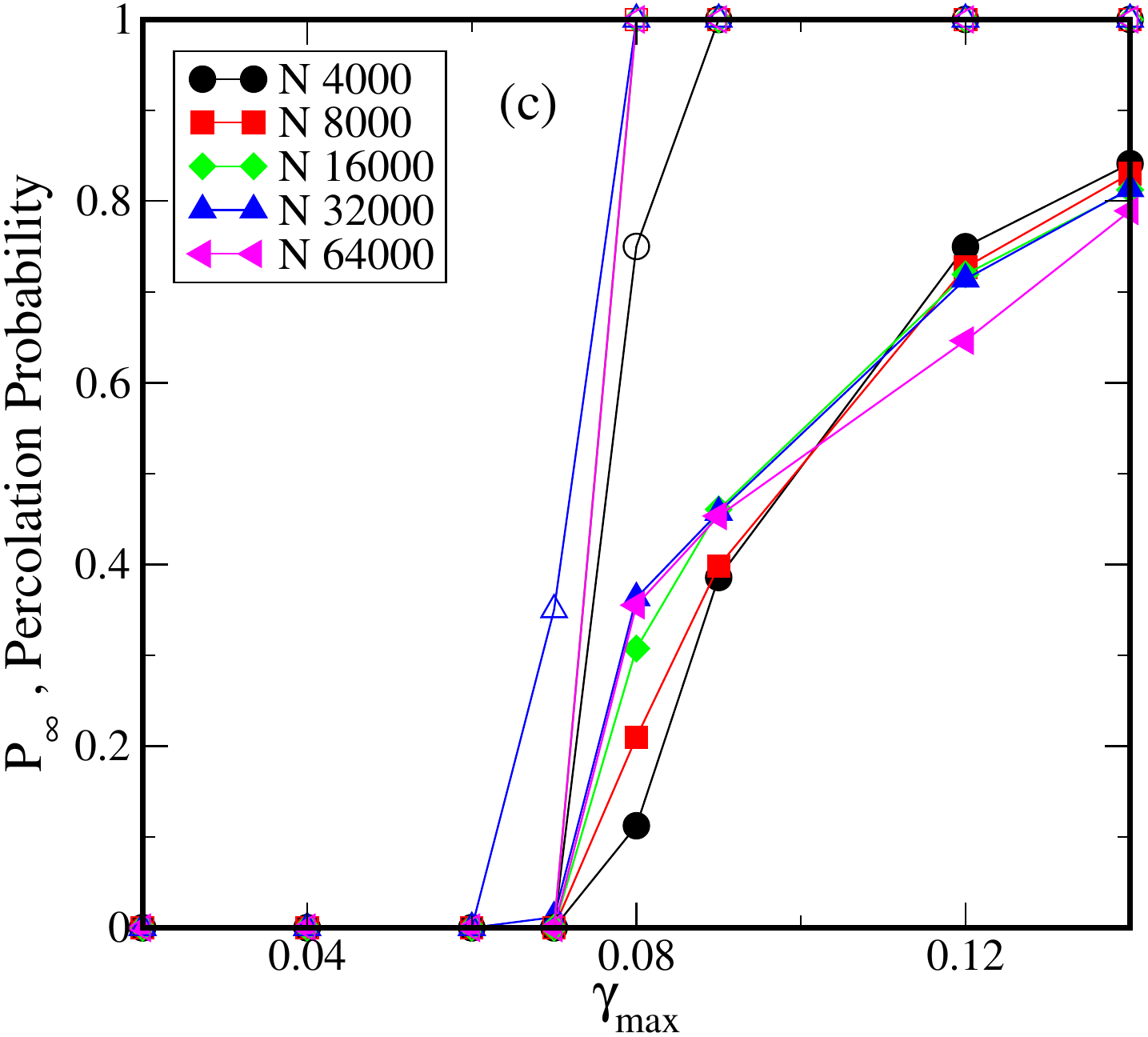}
\includegraphics[width=.24\textwidth]{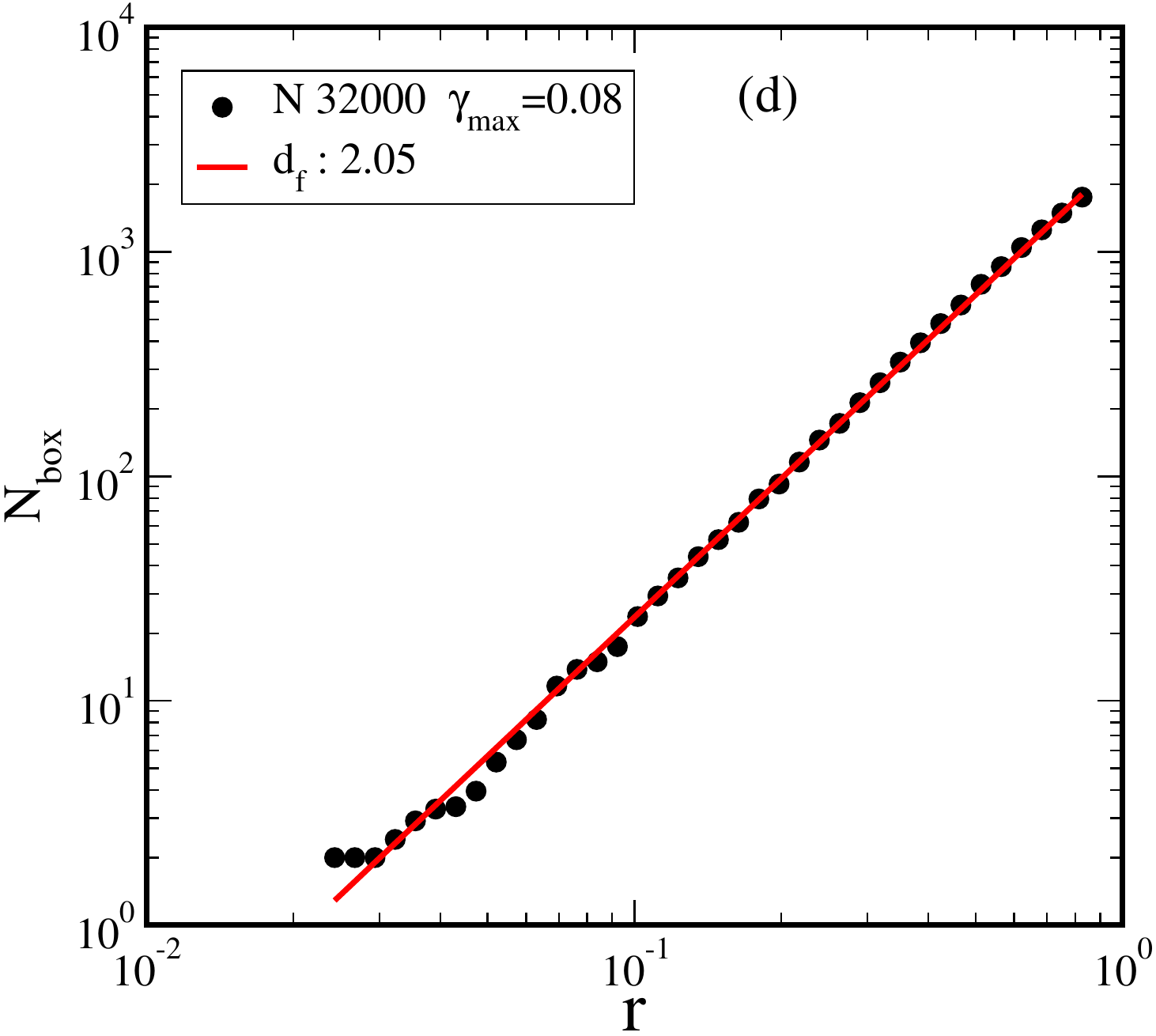}
\caption{ Percolation of avalanches and fractal
  dimension of percolating clusters: {\bf (a)} Percolation
probability and weight of the spanning cluster $P_{\infty}$ shown as
open and filled symbols respectively against the occupation number $P$
for different system sizes, considering all events, for $\gamma_{max}
= 0.08$. A percolation transition takes place for $P \simeq 0.05$
although the threshold is system size dependent.  {\bf (b)}
Percolation probability (inset) and $P_{\infty}$ averaged over all
events, {\it vs.}  $\gamma_{max}$.  {\bf (c)} Percolation probability
  and $P_{\infty}$ for the cumulative set of particles rearranging
  over a cycle, shown as open and filled symbols respectively {\it
    vs.} $\gamma_{max}$, indicating a percolation transition at the
    yielding strain $\gamma_y$. $P_{\infty}$ just above the transition
    increases with system size.  {\bf (d)} Fractal dimension
    estimation from box counting. A log-log plot of the number of
    occupied boxes ($N_{box}$) is shown {\it vs.} the magnification
    $r$. The slope results in an estimated fractal dimension $d_f =
    2.05$.  Data shown are for $T = 1$, and averages are  over the first quadrant. 
 }
\end{figure*}

{\it Percolation.-}
Finally, we analyse the spatial structure of the avalanches
briefly, by studying (i) the percolation, and (ii) fractal dimension, of the avalanches.
Below $\gamma_y$, none of the avalanches percolate, whereas
above, a finite fraction does so. 
Figure 4(a) shows the weight of the spanning cluster  $P_{\infty}$, and percolation probability $PP$  averaged
over  bins in ``probability"  $P$, obtained from the fraction of displaced particles, 
(see Methods) for different system sizes for $\gamma_{max} = 0.08$, indicating a percolation transition at $P \gtrapprox  0.05$. 
However, the threshold is system size dependent, and thus merits further investigation. 
In Fig 4 (b),  $P_{\infty}$ and $PP$ averaged over all considered events  are shown as a function of $\gamma_{max}$. 
The percolation probability does not become $1$,  a result of considering all the drop events. 
To address this artefact  we analyse the cumulative set of all particles displaced  in any of the events. 
 The $P_{\infty}$ and $PP$ values shown in Figure 4 (c) 
indicate that above $\gamma_y$, this cumulative set always percolates and the weight $P_{\infty}$ 
is comparable for different system sizes. However, $P_{\infty}$ at the 
smallest $\gamma_{max}$ above $\gamma_y$ appears to increase with system size, suggesting a discontinuous 
change  across $\gamma_y$. The variation of $P$ with $\gamma_{max}$ in either method also shows an apparently discontinuous behaviour across $\gamma_y$ (see Supporting Information). 

{\it Fractal Dimension.-}
We compute the fractal dimension of the spanning clusters using the box counting method (see Methods). Figure 4 (d) shows a log-log plot of the occupied boxes 
{\it vs.} magnification $r$ (the largest $r$ corresponds to the smallest box size, 
of $1.1 \sigma_{AA}$) for $\gamma_{max} = 0.08$, $N = 32000$. We find a fractal dimension of $d_f = 2.05$, 
close to $2$, which appears consistent with the possibility that yield events are quasi-two dimensional.  However, based on the system size  dependence of the mean cluster size, the fractal dimension deduced is $d_f \sim 1$~\cite{Liu2016}, which is 
at odds with the result here, and requires further investigation for it to be properly understood. 


The results that we have discussed demonstrate that a sharp yielding transition is revealed through oscillatory deformation of model glasses. The character of the avalanches is qualitatively different across the transition, being localised below the transition, and becoming extended above. Contrary to theoretical expectations for uniform deformation, the mean size of the avalanches does not diverge upon approaching the yielding transition, and prompts theoretical investigation, including development of suitable elasto-plastic models, of yielding under oscillatory deformation\cite{Eran2014}. A signature of yielding is instead revealed by the progressive sluggishness of annealing behaviour as the transition is approached. Both the avalanche statistics and percolation characteristics suggest a discontinuous yielding transition, which may be consistent with the suggestion that yielding is a first order transition \cite{Itamar2016yielding}. Finally, our results reveal systematic, non-trivial annealing behaviour of the glasses near the yielding transition, which we believe are of relevance to thermomechanical processing of metallic glasses. 
In particular, processing near the yielding transition, both above and below, may lead to
significant change of properties, which may be utilised according to specific design goals.

{\it Methods.-}
The model system we study is the Kob-Andersen binary (80:20) mixtures
of Lennard Jones particles. The interaction potential is truncated 
at a cutoff distance of $r_{c \alpha\beta}=2.5 \sigma_{\alpha \beta} $ such that both the potential and the force smoothly
go to zero as given by
\begin{eqnarray}
V_{\alpha\beta}(r)&=&4 \epsilon_{\alpha\beta} \left[ \left( \frac{\sigma_{\alpha\beta}}{r} \right)^{12} - \left( \frac{\sigma_{\alpha\beta}}{r} \right)^{6} \right]\nonumber \\   
&& + 4 \epsilon_{\alpha\beta}\left[c_{0 \alpha\beta} + c_{2 \alpha\beta}\left(\frac{r}{\sigma_{\alpha\beta}}\right)^{2}\right], r_{\alpha\beta} < r_{c\alpha\beta} 
\label{eqn:KABMLJmodel}
\end{eqnarray}
where $\alpha,\beta \in \{A,B\}$ and the parameters
$\epsilon_{AB}/\epsilon_{AA}=1.5$, $\epsilon_{BB}/\epsilon_{AA}=0.5$,
$\sigma_{AB}/\sigma_{AA}=0.80$, $\sigma_{BB}/\sigma_{AA}=0.88$.
Energy and length are in the units of $\epsilon_{AA}$ and
$\sigma_{AA}$ respectively, and likewise, reduced units are used for other quantities. The correction terms $c_{0\alpha\beta},
c_{2\alpha\beta}$ are evaluated with the conditions that the potential
and its derivative at $r_{c\alpha\beta}$ must vanish at the cutoff. 

The initial liquid samples are equilibrated at two temperatures, T =
0.466 and T = 1 using the Nos\'{e} Hoover thermostat, at reduced density $\rho = 1.2$.  Independent
samples are generated for each temperature and system size by further
evolving the equilibrated liquid configurations by performing the
molecular dynamics simulations of constant energy, which are
separated by the structural relaxation time ($\tau_{\alpha}$) obtained
from the self intermediate scattering function ($F_{s}(k,t)$). For the
uniform shearing data, we have atleast 100 samples for all the system
sizes. The avalanche data shown for cyclic shearing are for at least 20
samples for N $\leq$ 32000, and  10 samples for larger systems.  All the simulations
are carried out using LAMMPS \cite{Plimpton1995}.

Shear deformation of the model amorphous solids is done employing
athermal-quasi static (AQS) simulations which consist of two steps. An
affine transformation of coordinates $x^{\prime}=x+d\gamma \times z;~~
y^{\prime}=y;~~z^{\prime}=z$ is imposed, subsequently followed by an
energy minimization using the conjugate-gradient method with
Lees-Edwards periodic boundary conditions.  Strain steps of
$d\gamma=2\times10^{-4}$ are used throughout, except for $N=256000$
for which $d\gamma=5\times10^{-4}$.  Initial configurations are the
inherent structures (local energy minima) of equilibrated liquid
samples.  Potential energy and mean square displacements  are computed at $\gamma=0$ as functions of cycles to
 ascertain that steady states are reached, wherein the coordinates of particles, and properties
such as the potential energy $U$ and shear stress $\sigma_{xz}$ remain  (below yield strain)
unchanged at the end of each cycle, or  (above yield strain) become statistically unchanged
upon straining further, and exhibit diffusive motion as a function of the number of cycles. 
Steady states for strain
amplitudes of $\gamma_{max}=0.02,~0.04,~0.06,~0.07,~0.08
~0.09,~0.12,~0.14$ are studied for system sizes $N = 2000, 4000, 8000,
16000, 32000$ and $64000$. To further probe finite size effects, we
have consider amplitude below the yield transition at
$\gamma_{max}=0.04$ for $N = 128000$ and $256000$, and
$\gamma_{max}=0.14$ for $N = 128000$.

In the steady state, we compute the potential energy per particle and
stress for each strain step. Plastic events result in discontinuous
energy and stress drops. A parameter $\kappa = \frac{\delta
  U}{Nd\gamma^2}$ \cite{Lerner2009} exceeding a value of $100$ is used
to identify plastic events, where $\delta U$ is the change in energy
during minimisation after  a strain step. Avalanche sizes based on the magnitude of energy
drops and the cluster sizes of ``active'' particles (that undergo plastic displacements) are both computed.
Particles are considered active if they are displaced by more than
$0.1 \sigma_{AA}$. The choice of this cutoff is based on considering
the distribution of single particle displacements $\delta r$, which
are expected to vary as a power law $P(\delta r) \sim \delta
r^{-5/2}$ for elastic displacements around a plastic core, but
display an exponential tail corresponding to plastic rearrangements
(see, {\it e. g.}, \cite{Fiocco2013}). The separation is clear cut
only for small $\gamma_{max}$, and we choose the smallest cutoff value
(observed for $\gamma_{max} = 0.02$) so that plastic rearrangements at
all $\gamma_{max}$ are considered. In performing cluster analysis, two
active particles are considered to belong to the same cluster if they
are separated by less than $1.4 \sigma_{AA}$ (first coordination
shell).  The normalised histogram of cluster sizes $P(s)$ is obtained
from statistics for all the events. The mean cluster size is computed
from the distributions as $<s> = {\sum_s s^2 P(s) \over \sum_s s
  P(s)}$ (see Supporting Information).
  
 For the percolation analysis,  we consider all the plastic events in the 
first quadrant of the cycle ($\gamma$ from $0$ to $\gamma_{max}$),  and compute 
the ``probability'' $P$  from the fraction of particles that undergo plastic displacement, and 
the weight of the spanning cluster
$P_{\infty}$, from the fraction of particles that belong to the spanning cluster ($P_{\infty} = 0$ 
if there is no spanning cluster). The percolation probability $PP = 1$ if a spanning cluster is present and $0$ otherwise. 

To obtain the fractal dimension of percolating clusters, we employ the
method of box counting. The simulation volume is divided into boxes of
a specified mesh size, and the number of boxes that contain a part of
the cluster, $N_{box}$, is counted. This is repeated for a series of
mesh sizes, and the fractal dimension is obtained as the slope
$d_f=\frac{\log(N_{box})}{\log r}$ where $r$ is the inverse of mesh
size.

\begin{acknowledgments} We wish to thank J. L. Barrat,  P. Chaudhuri, M. Falk, G. Foffi,  A. L. Greer, J. Horbach, I. Procaccia, 
 M. Robbins,  A. Rosso and M. Wyart for useful discussions. We wish to specially thank H. A. Vinutha for discussions and help regarding computations reported here. 
We gratefully acknowledge TUE-CMS and SSL, JNCASR, Bengaluru for computational resources and support.  
\end{acknowledgments}

\bibliography{mybib}{}

\clearpage

\part*{}   

\setcounter{figure}{0}
\renewcommand{\thefigure}{S\arabic{figure}}%

\begin{center}
\textbf{The yielding transition in amorphous solids under oscillatory shear deformation (Supplementary Information)}
\\
{Premkumar Leishangthem, Anshul D. S. Parmar and Srikanth Sastry}
\date{\today}
\end{center}

Here we provide additional information regarding the following aspects of analysis of model glasses subjected to  oscillatory deformation: 
(i) Number of cycles required  to reach the steady state,
(ii) Evolution of energy with cycles, 
(iii) Identification of particles undergoing plastic displacements, 
(iv) Avalanche distributions at different strain for different strain amplitudes 
(v) Energy drops and avalanche sizes for uniform shear, and 
(vi) Probability of plastic displacement as a function of strain amplitude.

\section{Number of cycles required to reach the steady state}
\begin{figure}[htbp!] 
\centering 
\includegraphics[width=.305\textwidth]{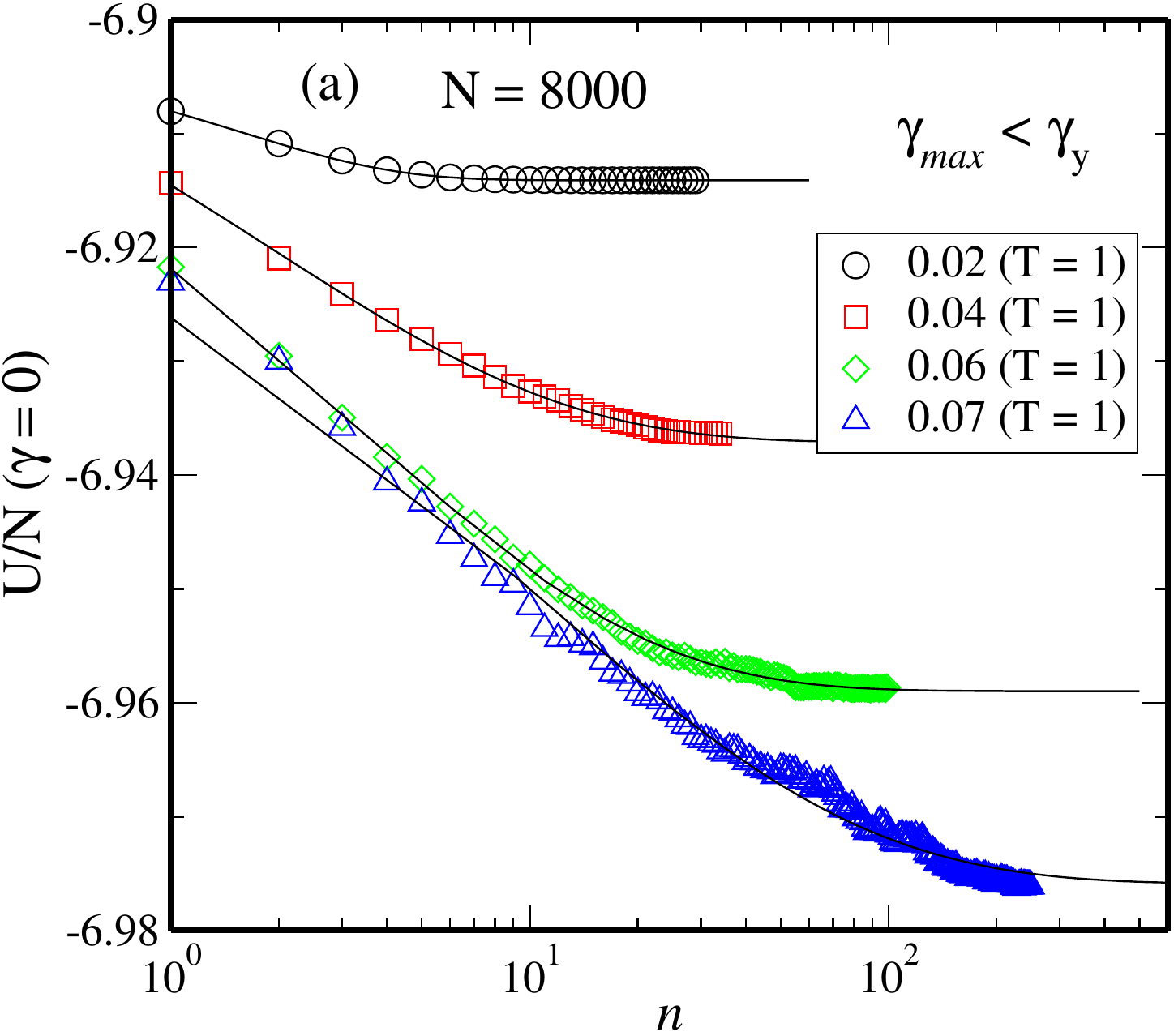}
\includegraphics[width=.303\textwidth]{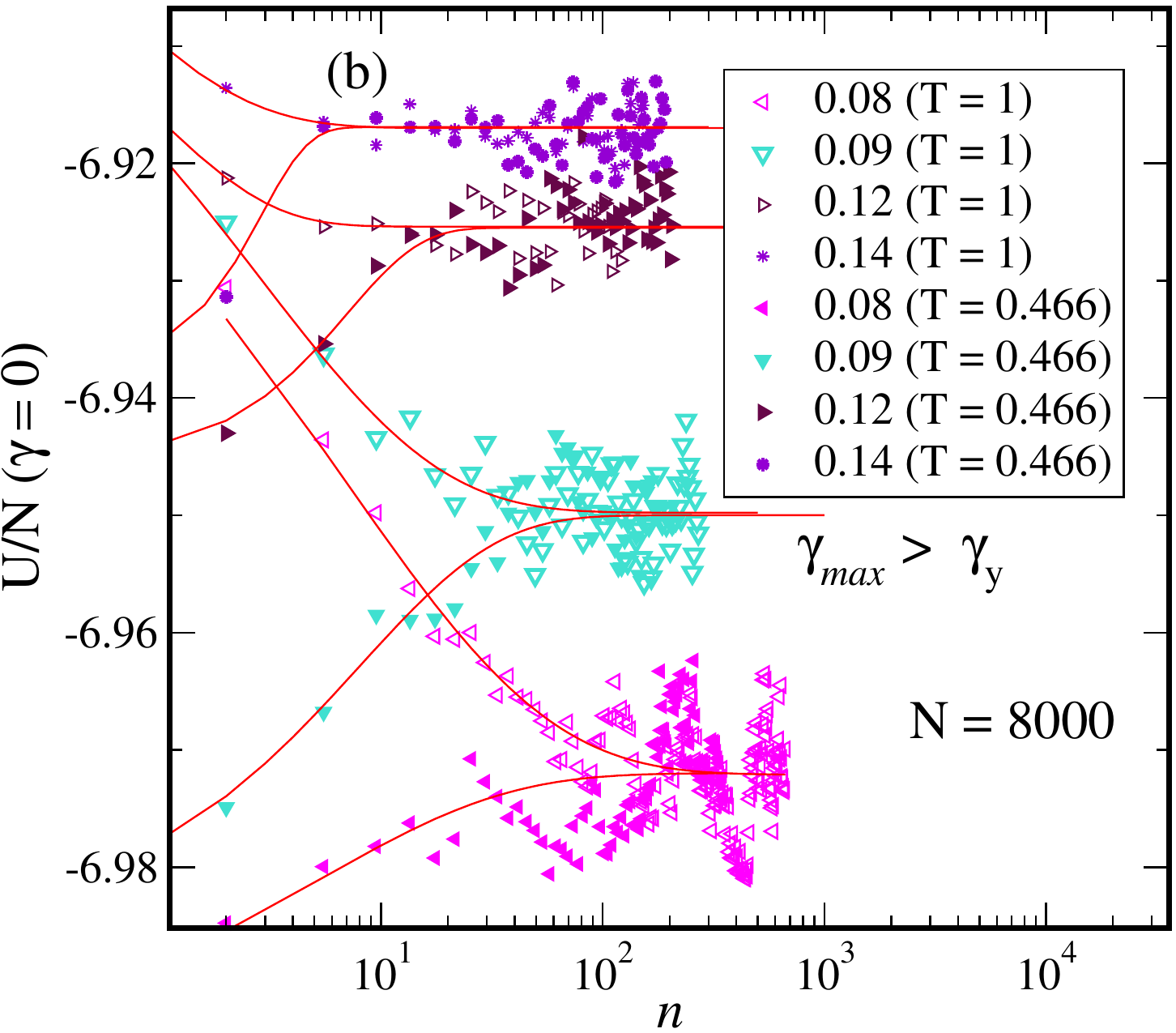}
\includegraphics[width=.291\textwidth]{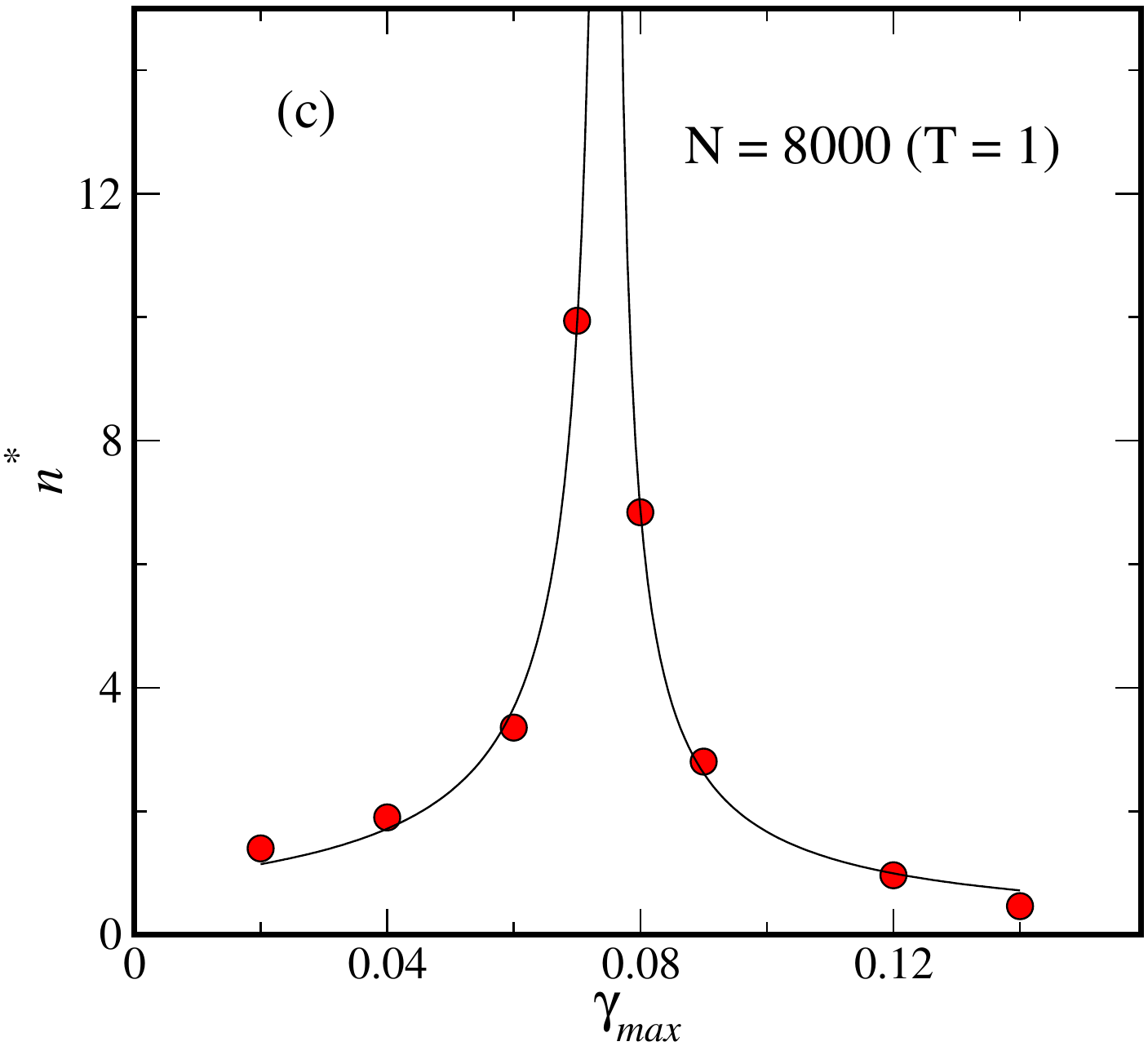}
\caption{Energy {\it vs.} cycle for different strain amplitudes (a) below
and (b) above the transition. (c) The decay constant $n^{*}$ to reach the steady state.}
\label{Fig:cyclic-nstar}
\end{figure}

We analyse the potential energy $U(\gamma=0)$ {\it vs.} the number of cycles $n$ 
to probe the approach to the steady state, and denote by $n^*$ the indicative 
number of cycles needed to reach the steady state, and is obtained as a decay constant by fitting the cycle dependence of energies to the form 
$U(n) = U_{\infty} + \Delta U_0 \exp\left[-(n/n^{*})^\beta\right]$. These fits are shown in Fig. \ref{Fig:cyclic-nstar} (a), (b) for $\gamma_{max}<\gamma_y$ (for the $T = 1$ cases) 
and $\gamma_{max}>\gamma_y$ (for both $T = 1$ and $T = 0.466$) respectively. Fig. \ref{Fig:cyclic-nstar}(c) shows that $n^{*}$ grows strongly on approaching the yield value of $\gamma_{max}$
between $0.07$ and $0.08$. The fit lines are guides to the eye. 
 
\clearpage

\section {Evolution of energy with cycles} 
\begin{figure}[htbp!] 
\centering 
\includegraphics[width=.44\textwidth]{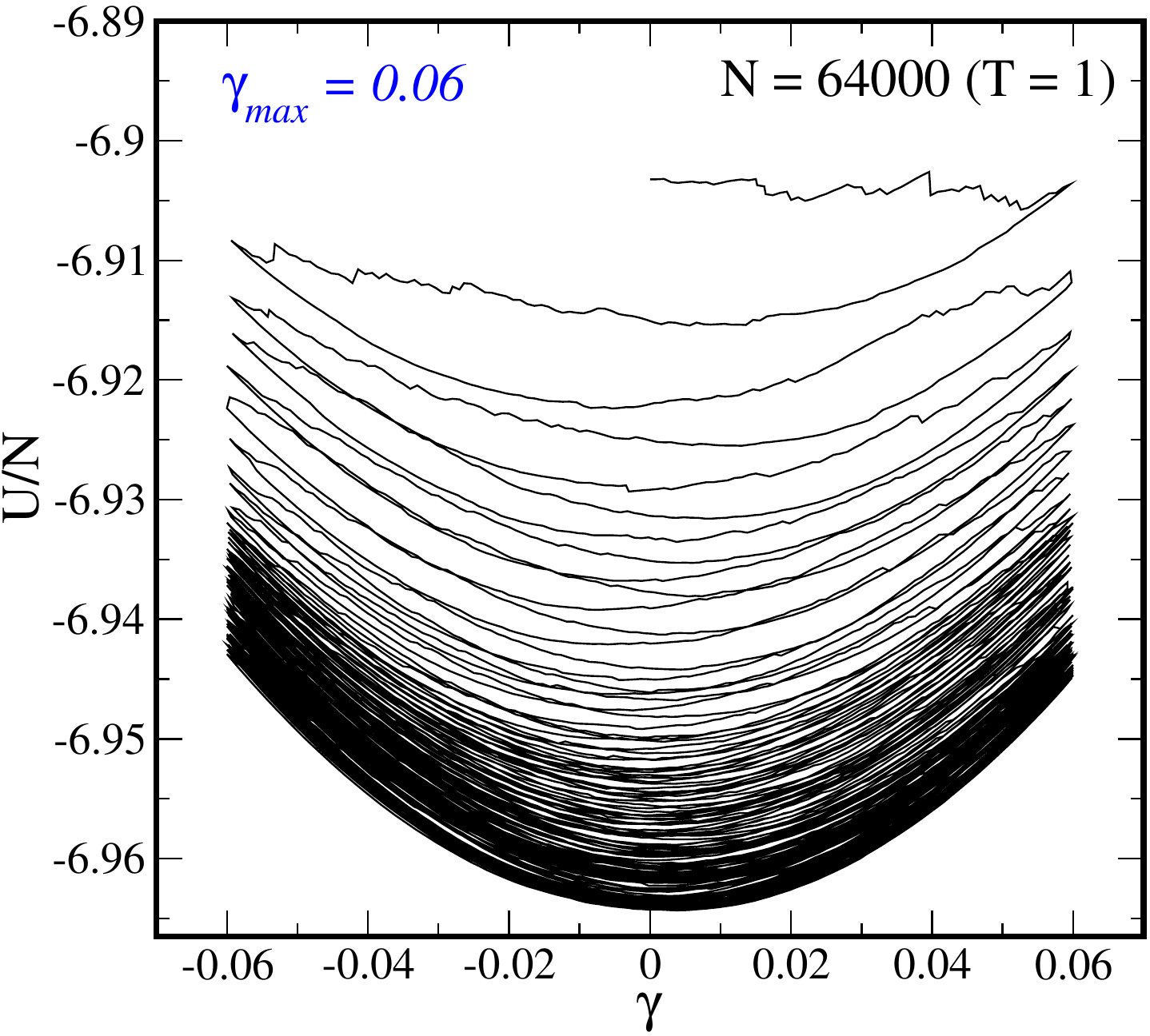}
\includegraphics[width=.44\textwidth]{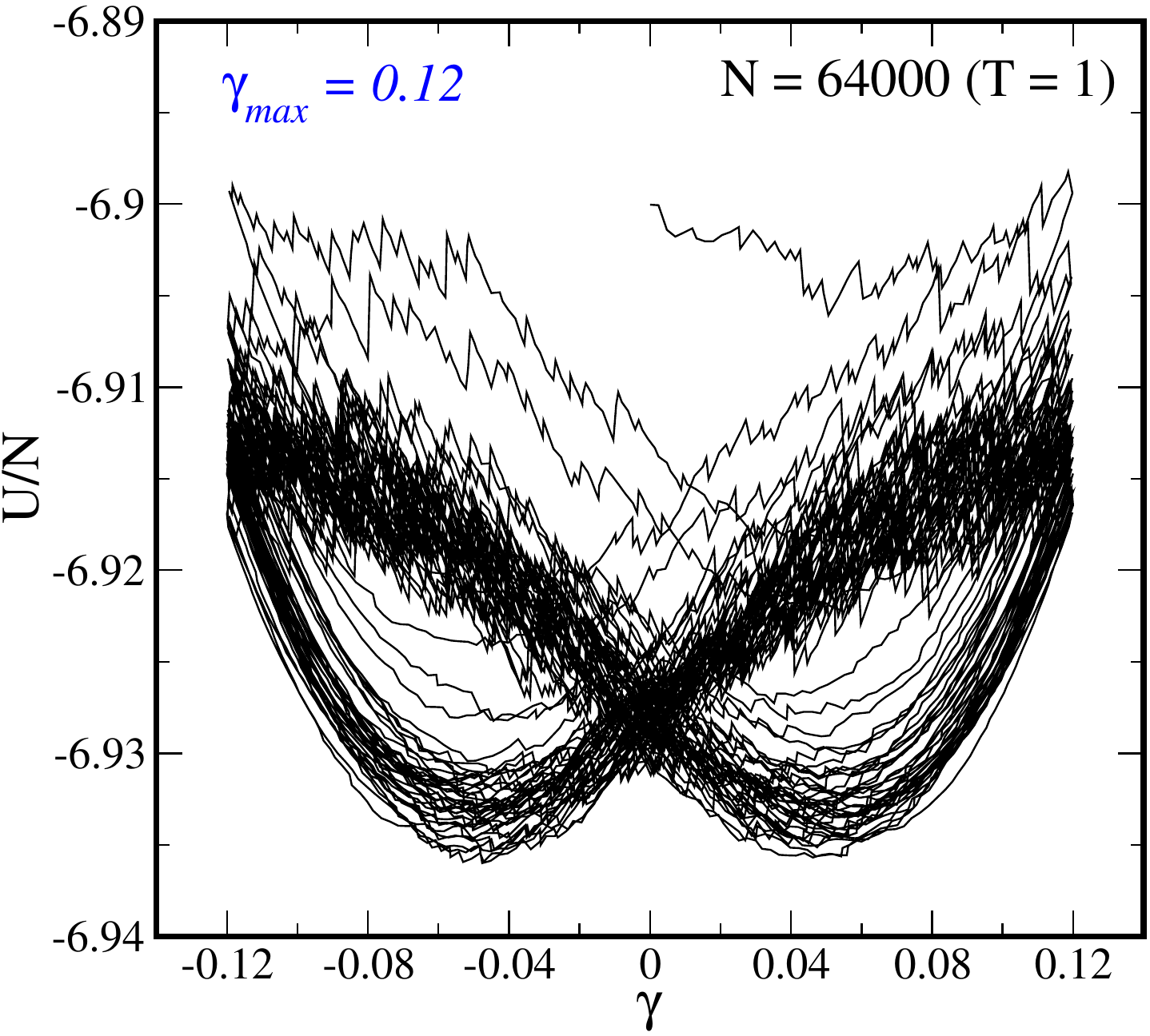}
\caption{Energy {\it vs.} strain over cycles starting with the undeformed glass, showing (for $\gamma_{max} = 0.06$) the approach to a single minimum at $\gamma = 0$, and bifurcation into two minima (for $\gamma_{max} = 0.12$) at finite strain.}
\label{Fig:batman}
\end{figure}

\begin{figure}[htbp!] 
\centering 
\includegraphics[width=.43\textwidth]{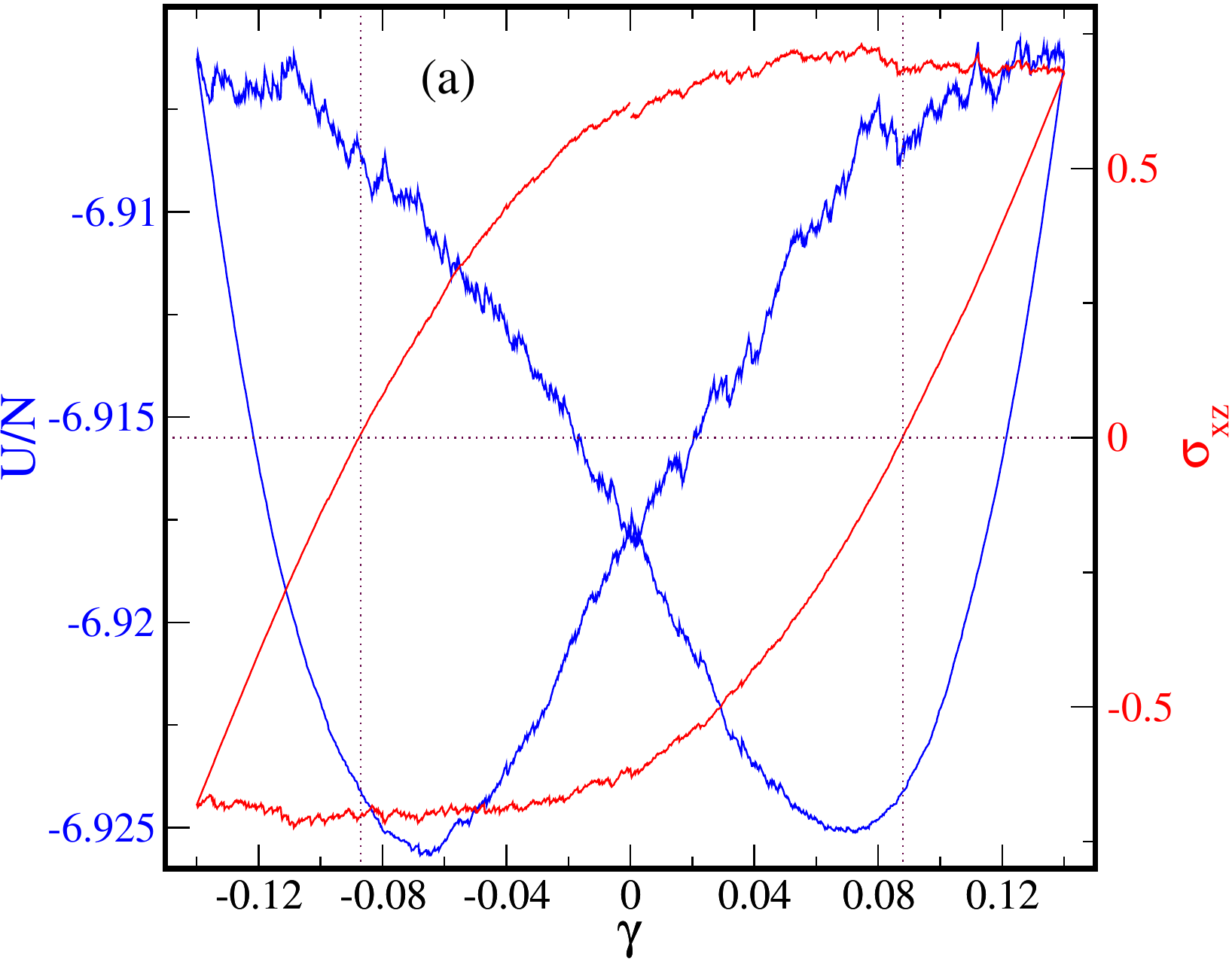}
\includegraphics[width=.43\textwidth]{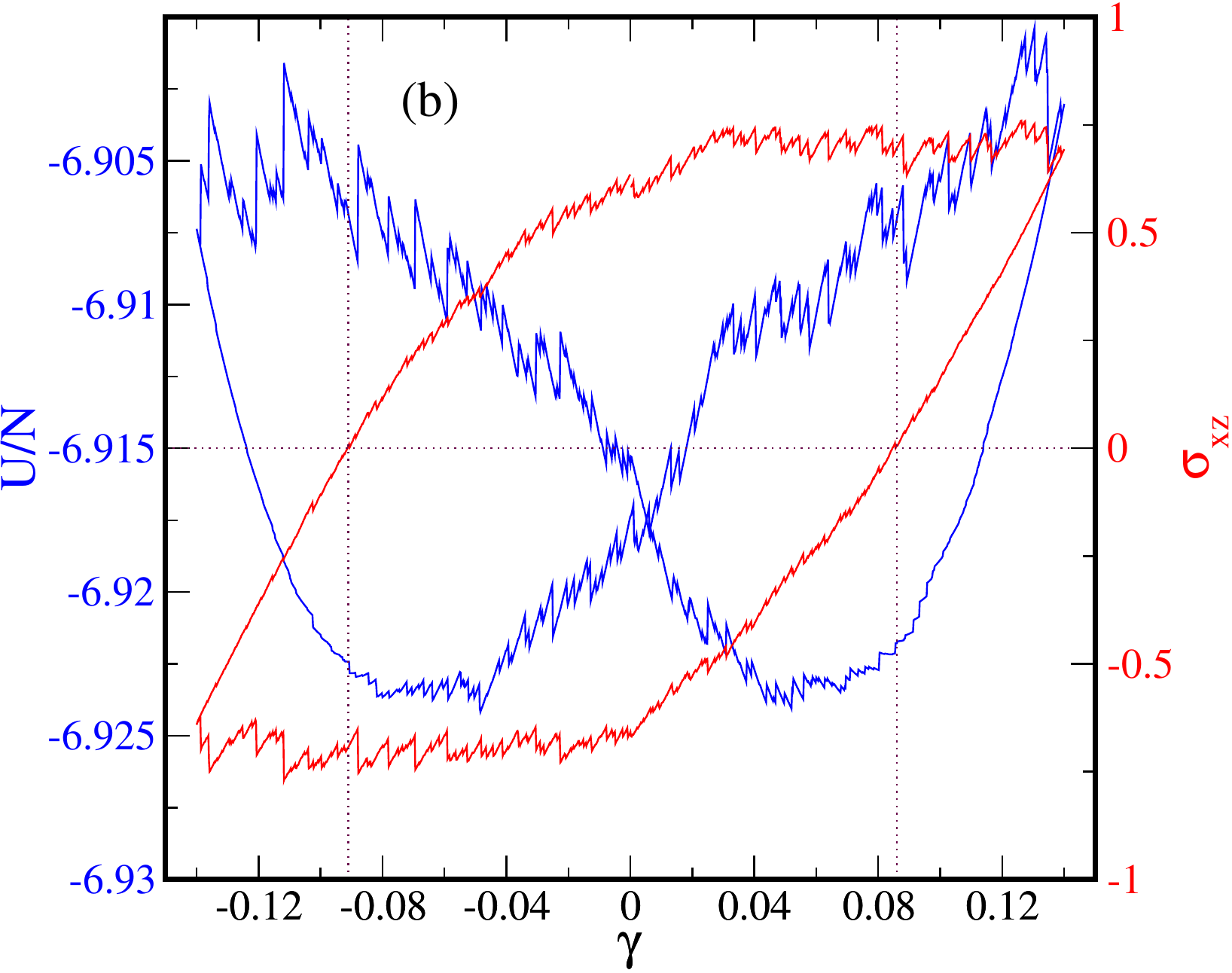}
\caption{ Stress {\it vs.} strain (red lines) and energy {\it vs.} strain (blue lines) are
shown, along with with dotted lines as guides to the eye, to locate $\sigma_{xz}$ = 0.
Data shown are (a) averaged over 10 cycles and (b) for a single cycle, in the steady state.}
\label{Fig:cyclic-steadystatesES}
\end{figure}

We present here the potential energy at each strain step 
over the cycles of deformation for $\gamma_{max}$ = $0.06$ and $0.12$ in 
Fig. \ref{Fig:batman}. The strain at $U=U_{min}$ and $\sigma_{xz} = 0$ are represented by 
$\gamma_{U_{min}}$ and  $\gamma_{\sigma_0}$. To test if the location of the energy minima coincide with the strain at zero stress, {\it i. e.} if 
$\gamma_{U_{min}}$ =  $\gamma_{\sigma_0}$, we plot the energy and the stress
loops in Fig. \ref{Fig:cyclic-steadystatesES} 
for $\gamma_{max}$ = 0.14 of N = 64000 (T=1). Though their values are close, we find  $\gamma_{U_{min}} \neq \gamma_{\sigma_0}$ 
(for $\gamma_{max}>\gamma_y$).

\section{Identification of particles undergoing plastic displacements}
\begin{figure}[htbp!] 
\centering 
\includegraphics[width=0.42\textwidth]{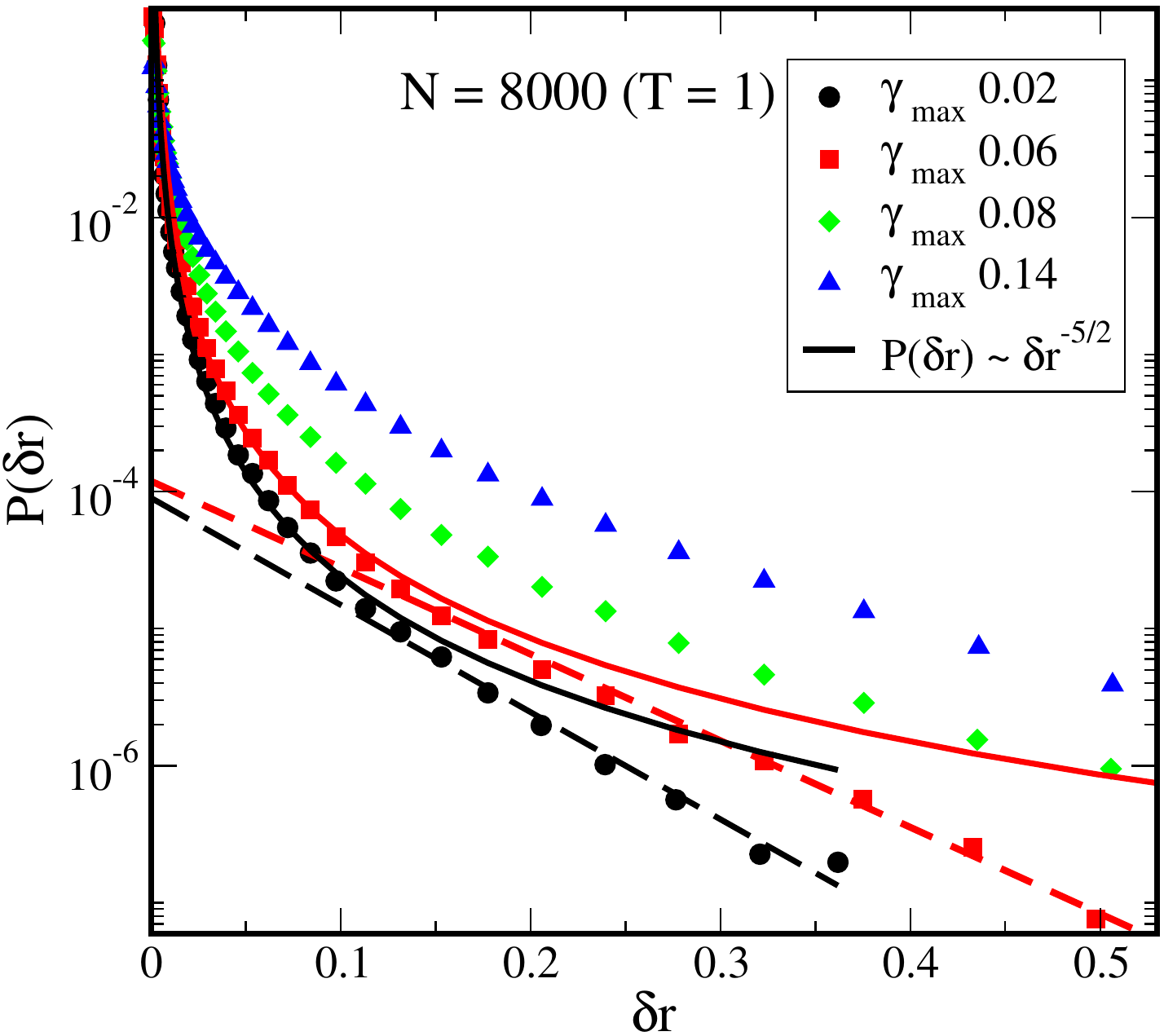}
\includegraphics[width=0.42\textwidth]{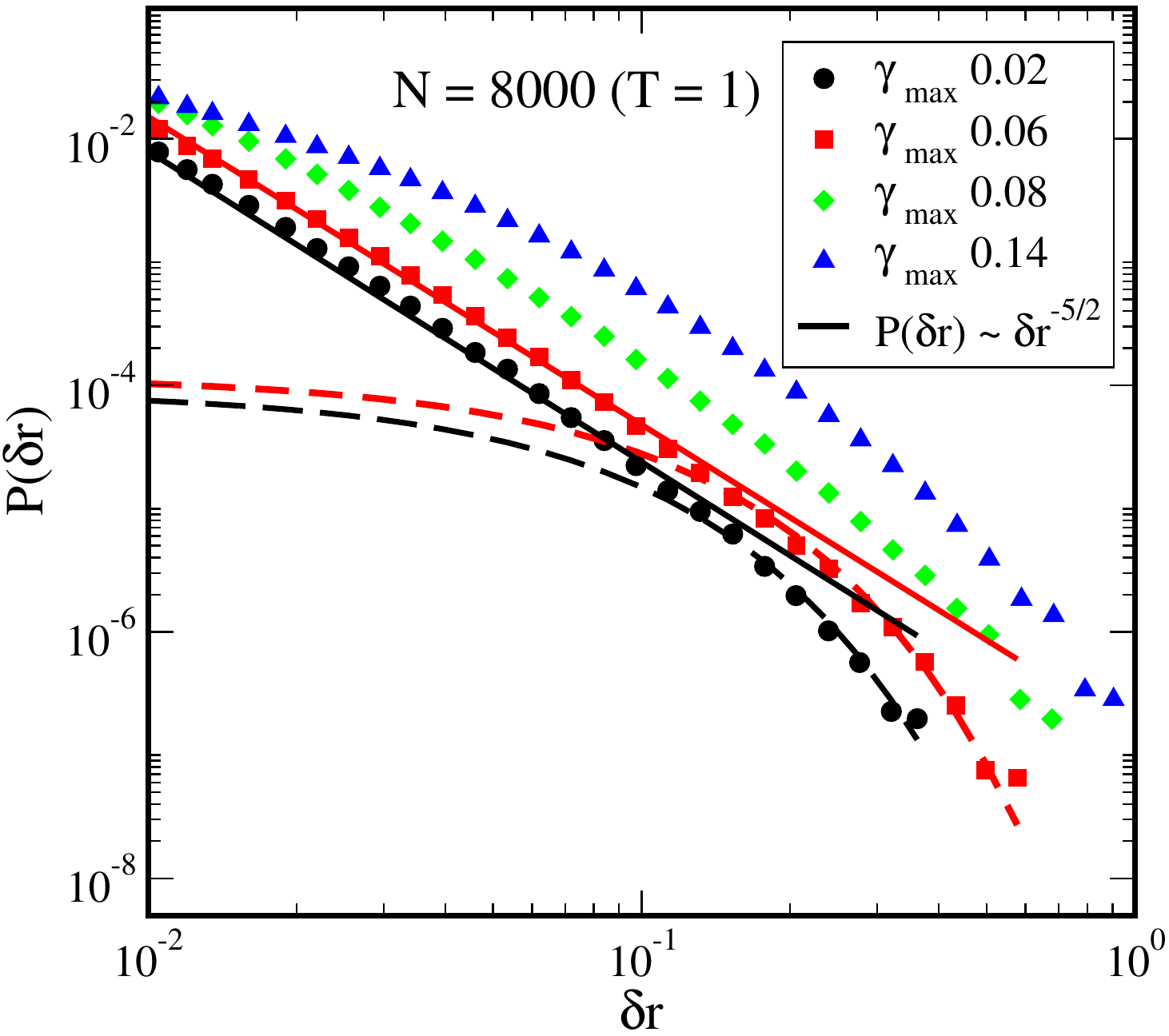}
\caption{Distributions of the particle displacements at the plastic
events shown in semi-log and log-log scales for various strain amplitudes.}
\label{Fig:delr}
\end{figure}

Here we describe how particles that are labeled ``active", that undergo plastic deformation during an energy drop, are identified, based on previous work [see, {\it e. g.} 
D. Fiocco, G. Foffi and S. Sastry, Phys. Rev. Lett. 025702 (2014); T. B. Schro{}der, S. Sastry, J. C. Dyre, and S. C. Glotzer, J.
Chem. Phys. 112, 9834 (2000)]. In the presence of a plastic rearrangement, it is found that the distribution of single particle displacements $p(\delta r)$ displays an exponential tail, corresponding to plastic displacements, and a power law distribution at smaller values with an exponent of ${-5/2}$ which may be deduced from assuming that the rest of the system undergoes an elastic deformation owing to the stresses created by the plastic deformation. As seen in Fig. \ref{Fig:delr}, such an expectation is clearly satisfied at low strain amplitudes $\gamma_{max}$, but (a) the location of the crossover depends on the strain amplitude, and (b) the distinction becomes less clear at large strain amplitudes. We wish to  include all particles that take part in plastic deformation, but to exclude those undergoing elastic displacements. As a conservative choice of cutoff, we use the cutoff 
that is clear and applicable for the case of $\gamma_{max} = 0.02$, namely  $\delta r = 0.1$.

\section{Avalanche distributions at different strain for different strain amplitudes}
\begin{figure}[htbp!] 
\centering 
\includegraphics[width=0.42\textwidth]{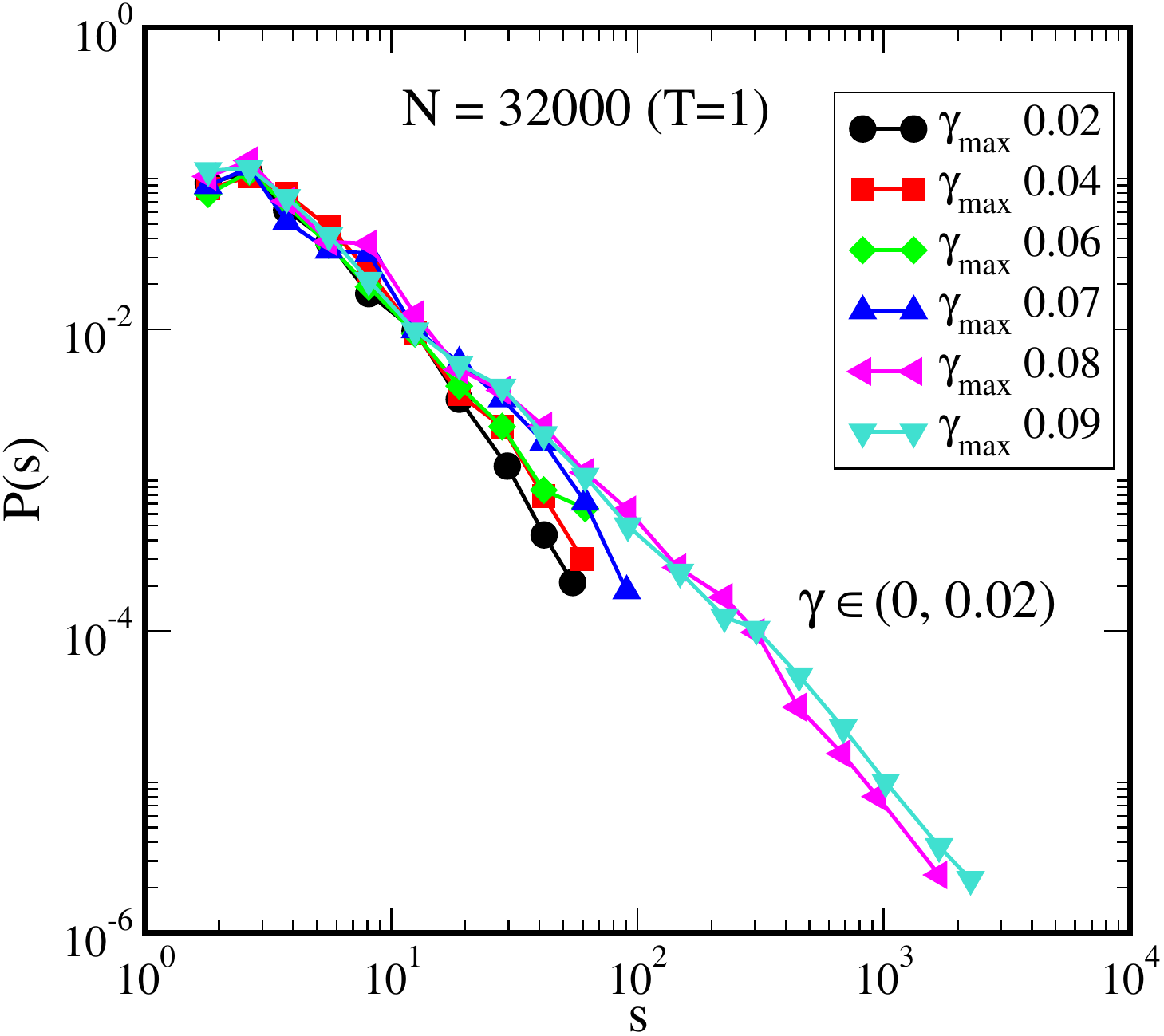}
\includegraphics[width=0.42\textwidth]{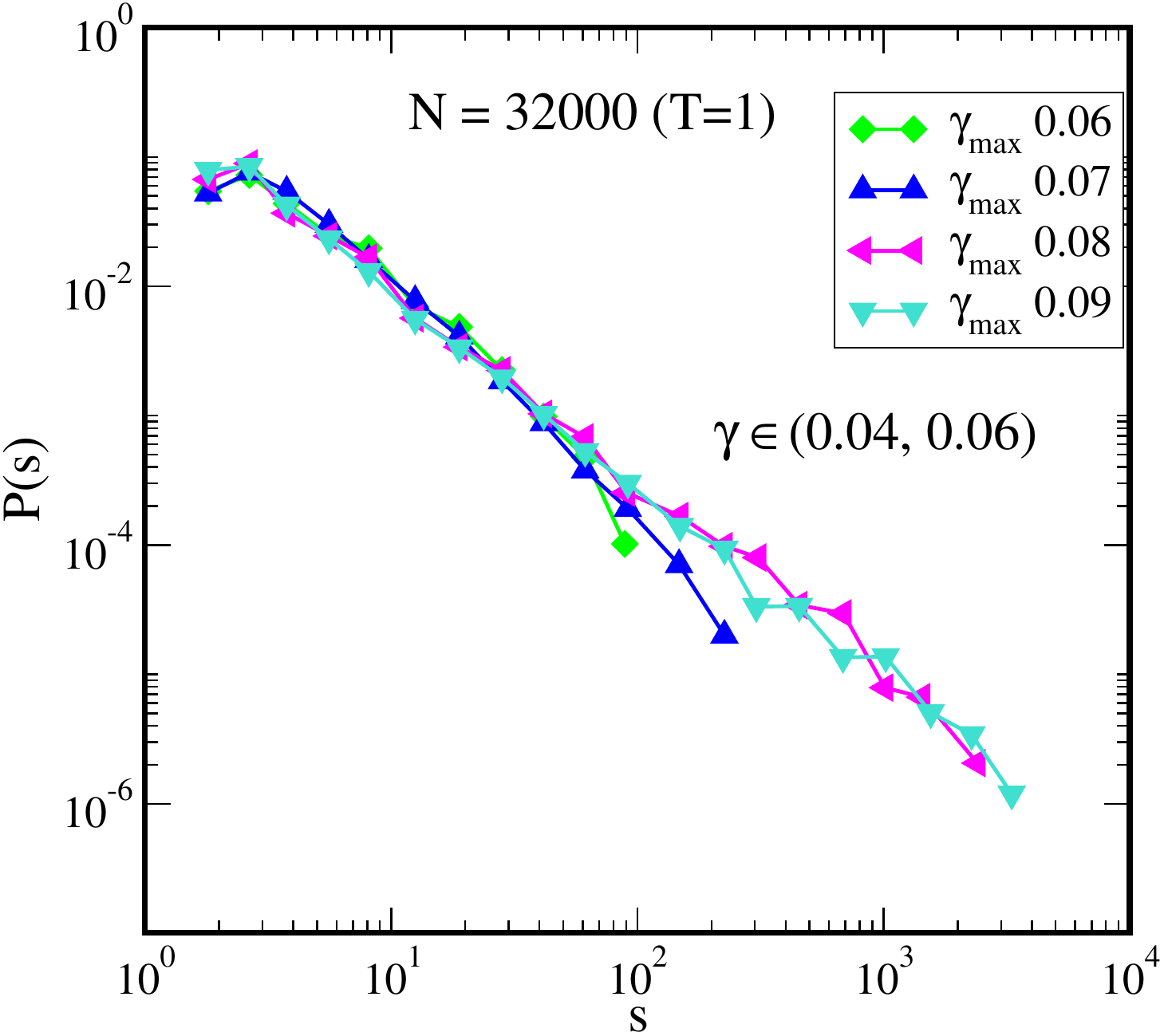}
\caption{Distributions of avalanche sizes within specified windows of strain [(left panel) $0$ to $0.02$, and (right panel)  $0.04$ to $0.06$] 
for different strain amplitudes.}
\label{Fig:s8}
\end{figure}

We show here the distribution of avalanche sizes that result when specific bins in the strain $\gamma$ are considered, for different amplitudes $\gamma_{max}$, with results averaged over the first quadrant of cycles of strain. 
Fig. \ref{Fig:s8} shows the distribution of avalanche sizes in the strain window of (a) $0$ to $0.02$, and (b)  $0.04$ to $0.06$, for different values of $\gamma_{max}$ for which we sample the strain window in the course of a full cycle. We note that in both bases, the distributions fall into two categories, one with $\gamma<\gamma_y$ and the other with $\gamma>\gamma_y$. In each category,  the distributions are largely independent of the value of  $\gamma_{max}$, but the distributions for the two categories are distinct. 

\section{Energy drops and avalanche sizes for uniform shear}
\begin{figure}[htbp!] 
\centering 
\includegraphics[width=0.43\textwidth]{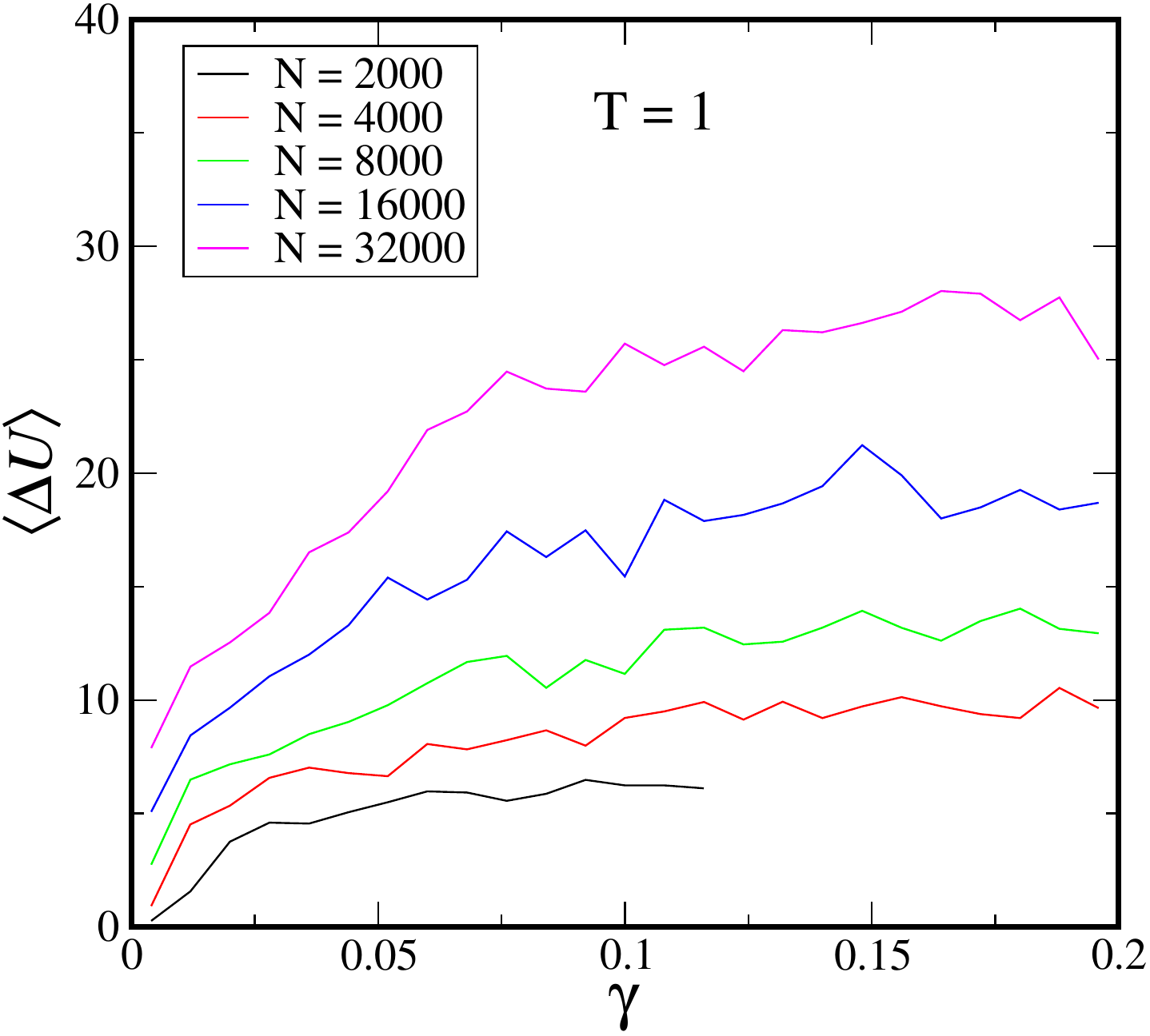}
\includegraphics[width=0.42\textwidth]{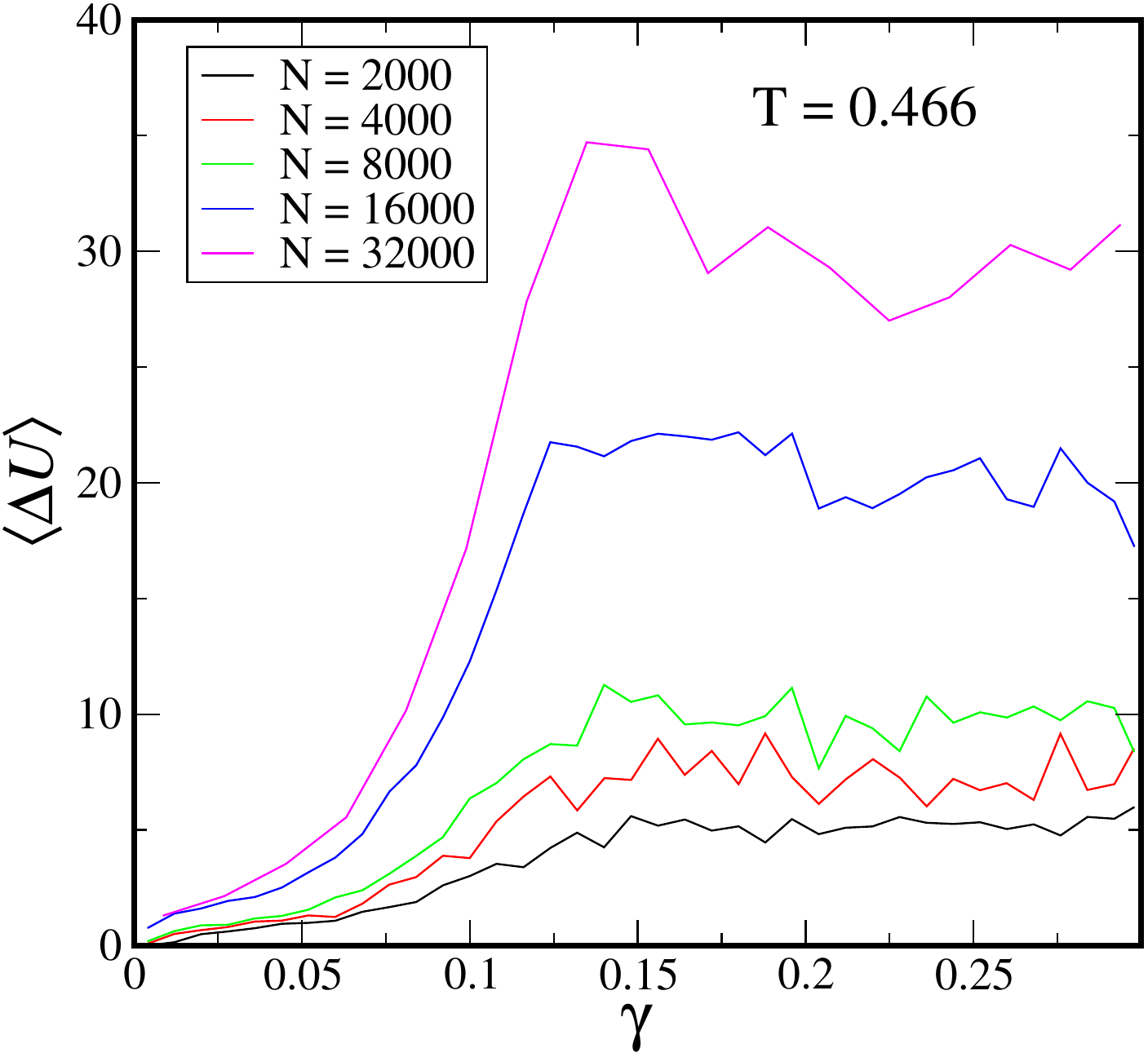}
\caption{Mean energy drops {\it vs.} strain for various system sizes for uniform strain.}
\label{Fig:uniform-mean-avalanche}
\end{figure}

\begin{figure}[htbp!] 
\centering 
\includegraphics[width=0.42\textwidth]{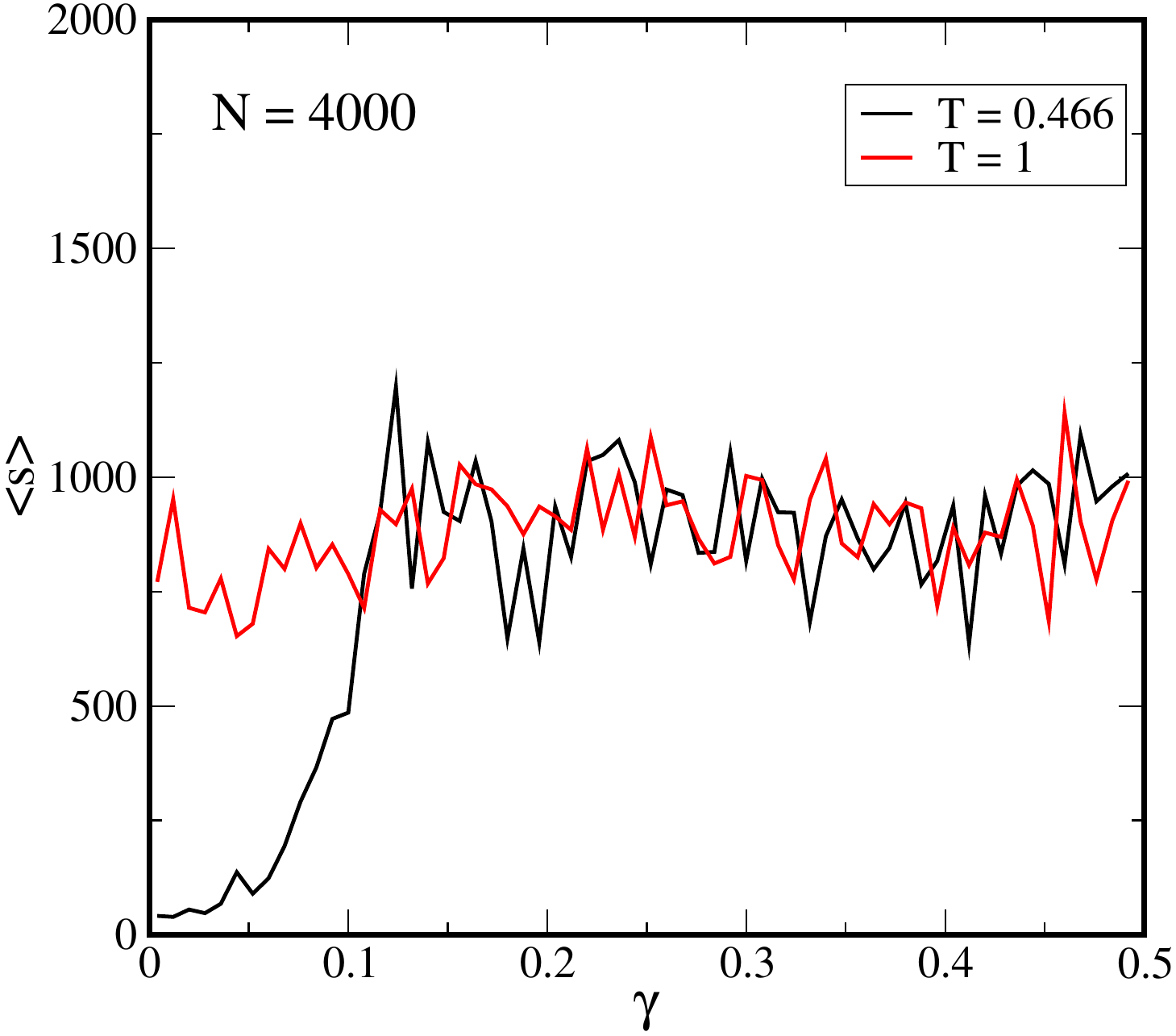}
\includegraphics[width=0.41\textwidth]{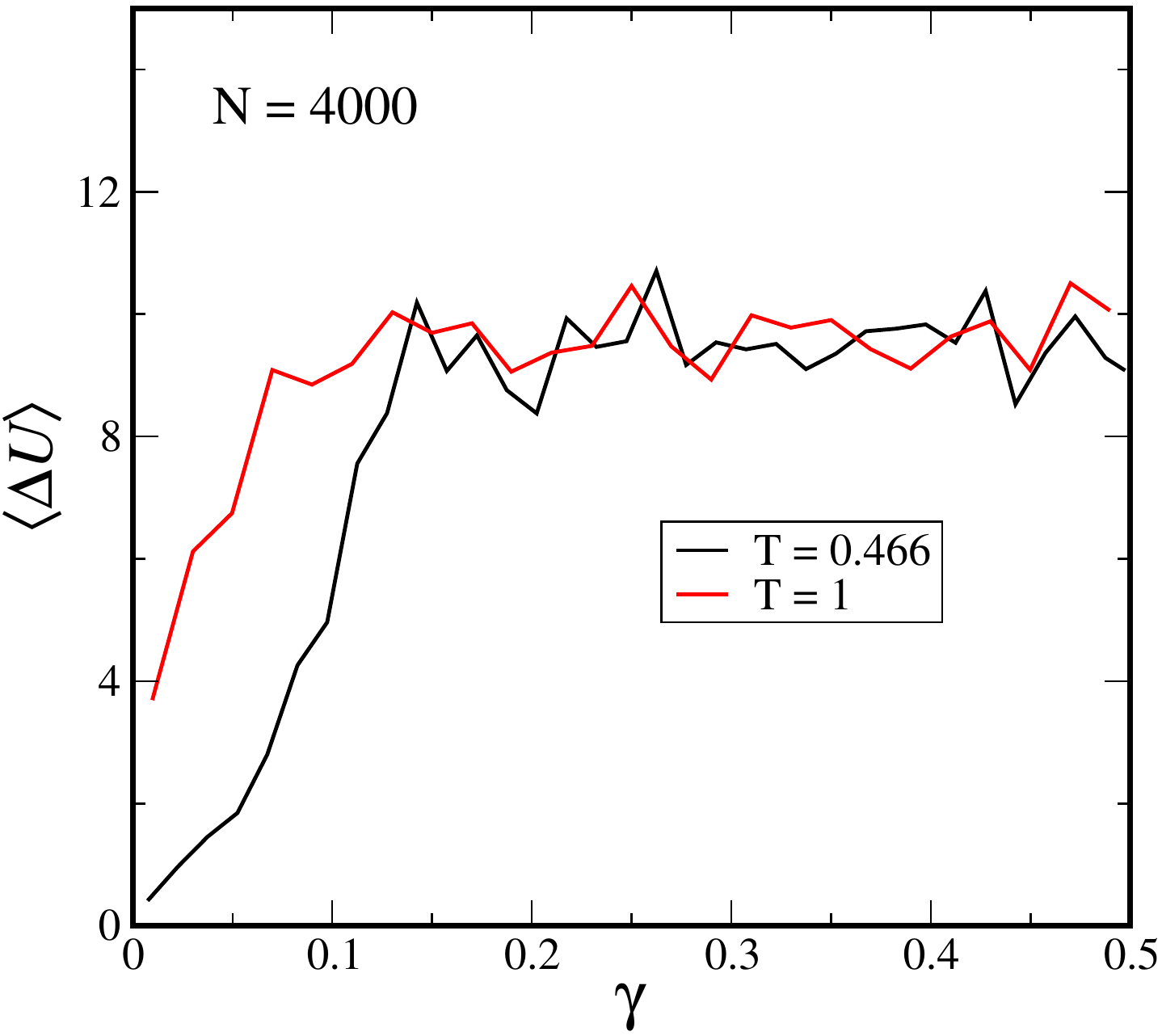}
\caption{Mean avalanche size and mean energy drops {\it vs.} strain for $T = 1$ and $T = 0.466$, $N = 4000$. }
\label{Fig:uniform-mean-compared}
\end{figure}

\begin{figure}[htbp!] 
\centering 
\includegraphics[width=0.42\textwidth]{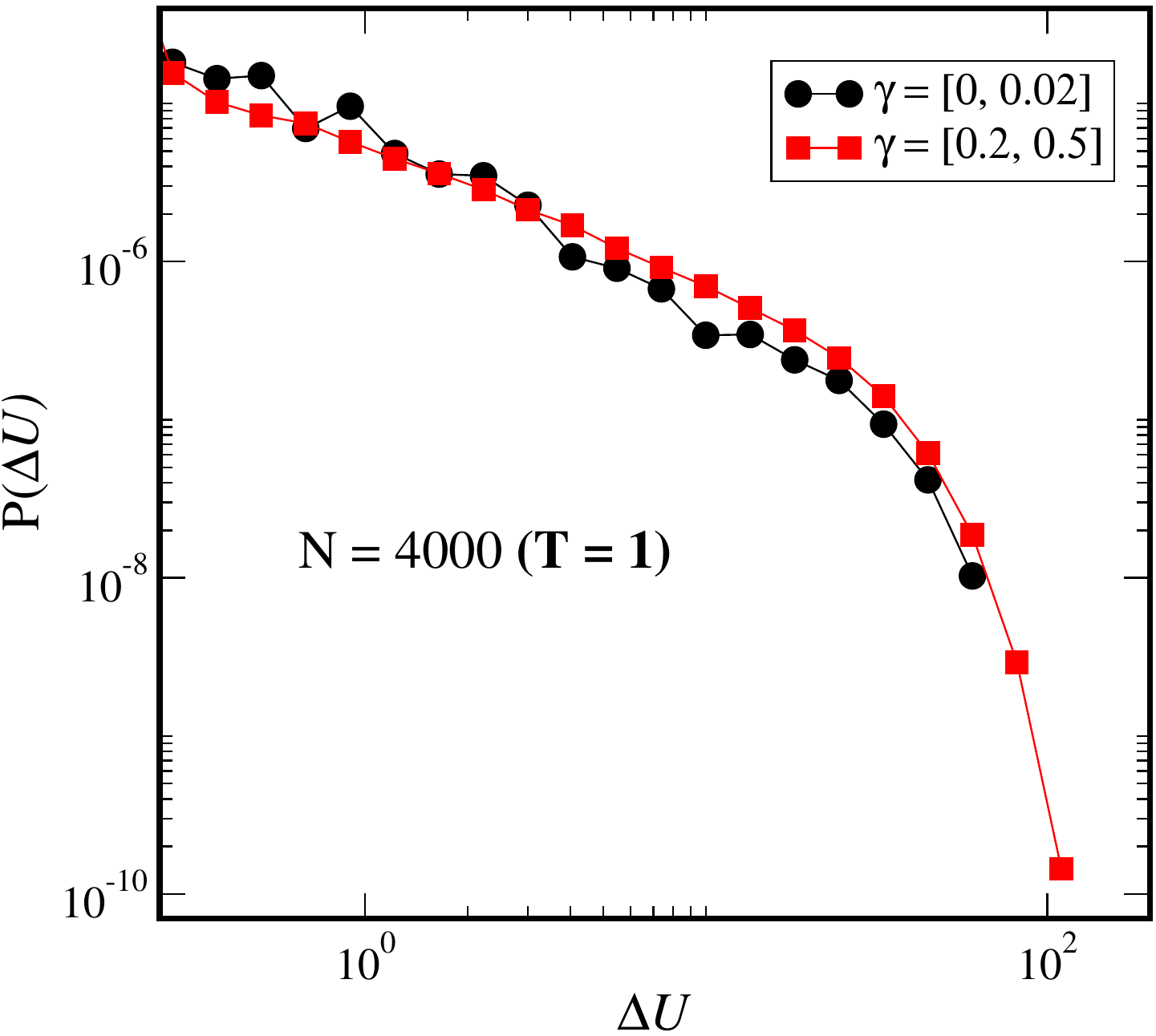}
\includegraphics[width=0.42\textwidth]{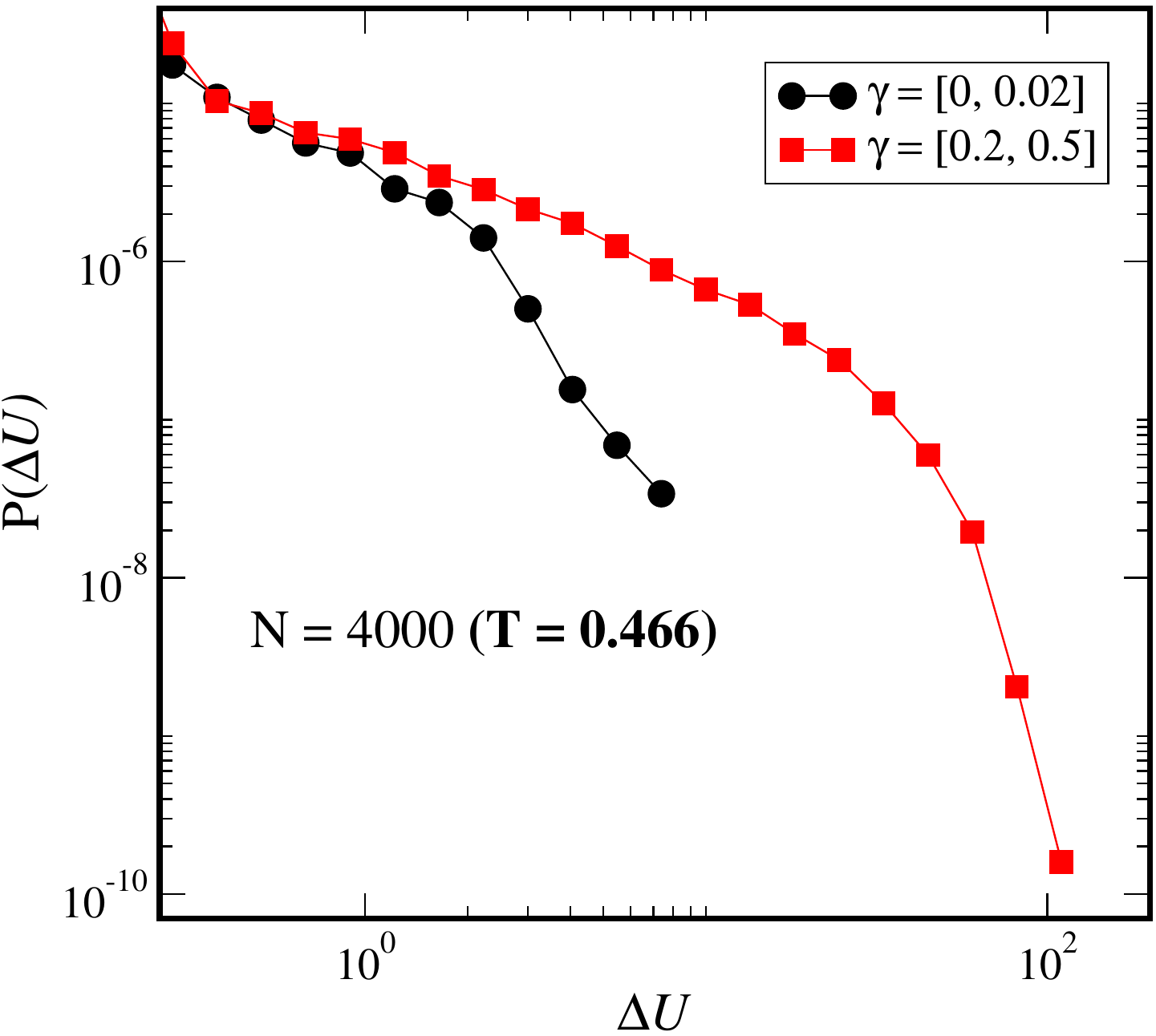}
\caption{Distributions of energy drops for two windows of strain, below and above the yielding transition, shown for $T = 1$ and $T = 0.466$.} 
\label{Fig:uniform-pdf-U}
\end{figure}

\begin{figure}[htbp!] 
\centering 
\includegraphics[width=0.42\textwidth]{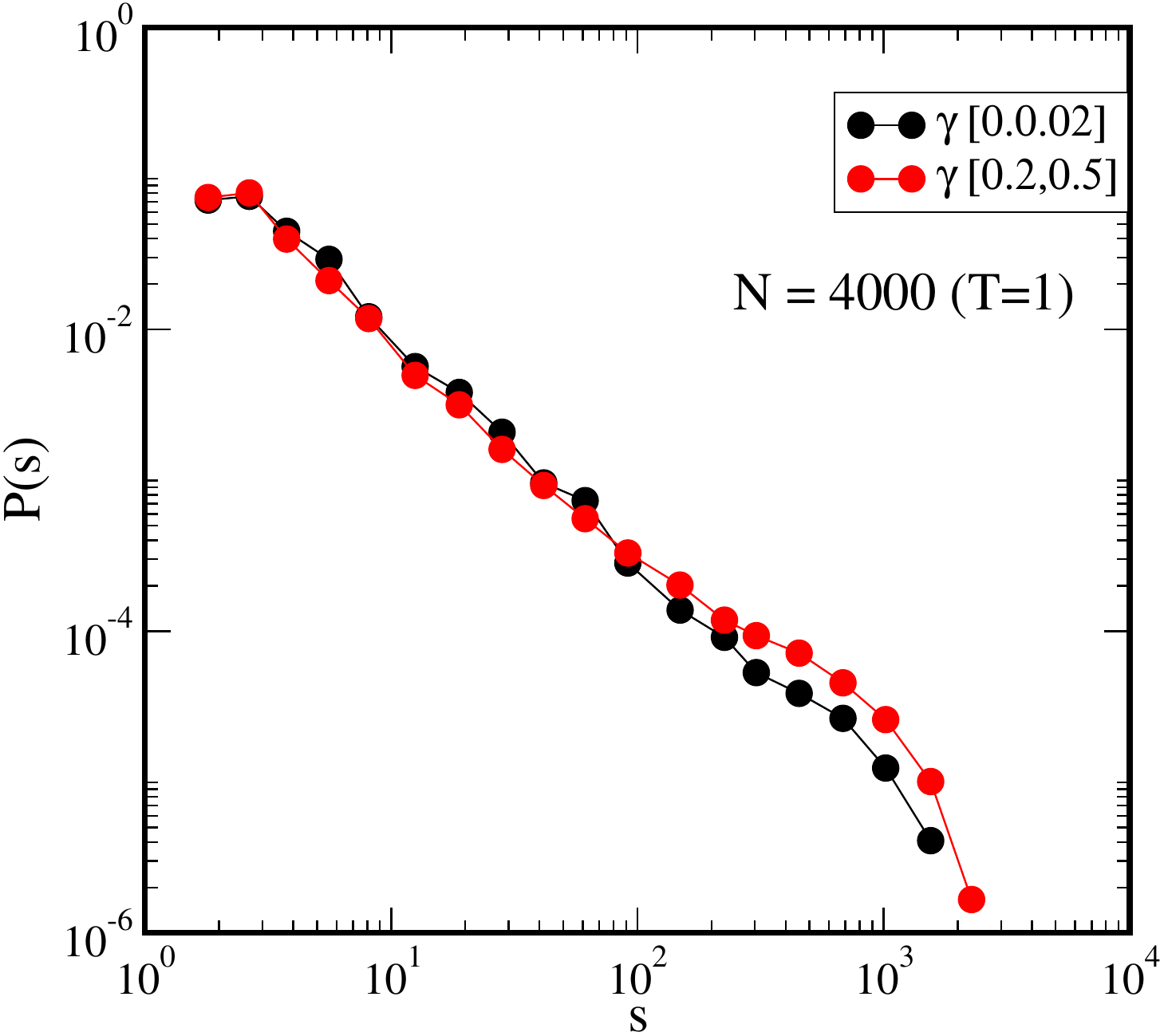}
\includegraphics[width=0.42\textwidth]{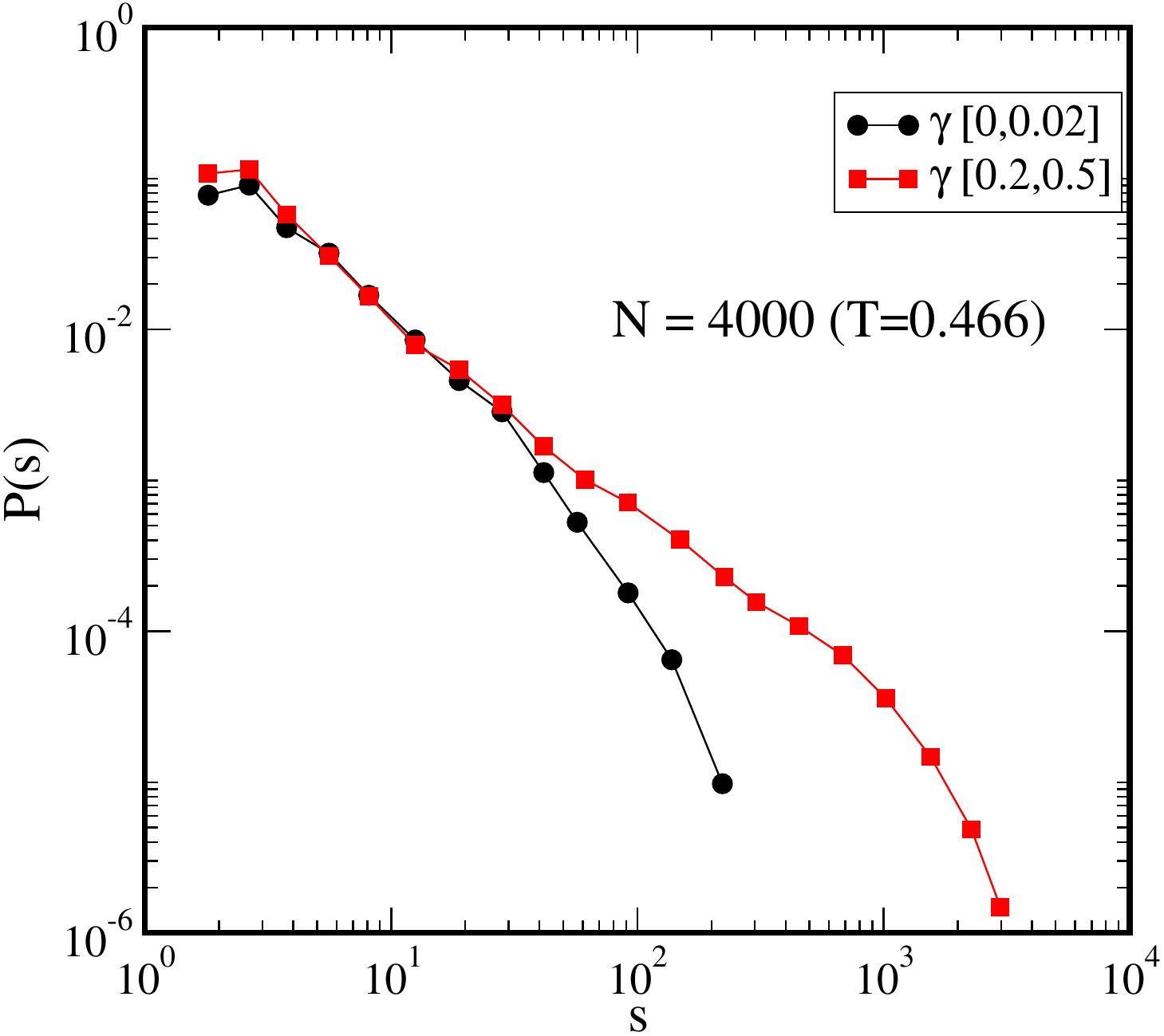}
\caption{Distributions of avalanche sizes for two windows of strain, below and above the yielding transition, shown for $T = 1$ and $T = 0.466$.}
\label{Fig:uniform-pdf-s}
\end{figure}

In Fig. \ref{Fig:uniform-mean-avalanche}  we present the mean energy drop  for uniform strain, for $T = 1$ and $T = 0.466$, for a range of system sizes for a range of system sizes. We see that the mean energy drops for the two temperatures are significantly different, but in each case show similar trends in their system size dependence. In Fig. \ref{Fig:uniform-mean-compared}, we compare, for $N = 4000$, the mean  size of avalanches and the mean energy drop, for $T = 1$ and  $T = 0.466$. The behaviour for the two cases is very different at strains below the yield strain, thus making it difficult to provide a general characterisation of the avalanches below the yield strain identified by oscillatory deformation. Further, we note that for $T = 1$, the energy drops and avalanche sizes below the yield strain remain high and comparable to values above yield strain.  
Fig. \ref{Fig:uniform-pdf-U} shows the distributions of energy drops for two strain intervals (one below, \{0, 0.02\},  and one above, \{0.2, 0.5\}, the yield strian). For $T = 1$, the two distributions do not differ, whereas for $T = 0.466$, they are widely separated. The avalanche size distributions shown in Fig. \ref{Fig:uniform-pdf-s} show the same pattern.

\clearpage


\section{Probability of plastic displacement as a function of strain amplitude}
\begin{figure}[htbp!] 
\centering 
\includegraphics[width=0.4\textwidth]{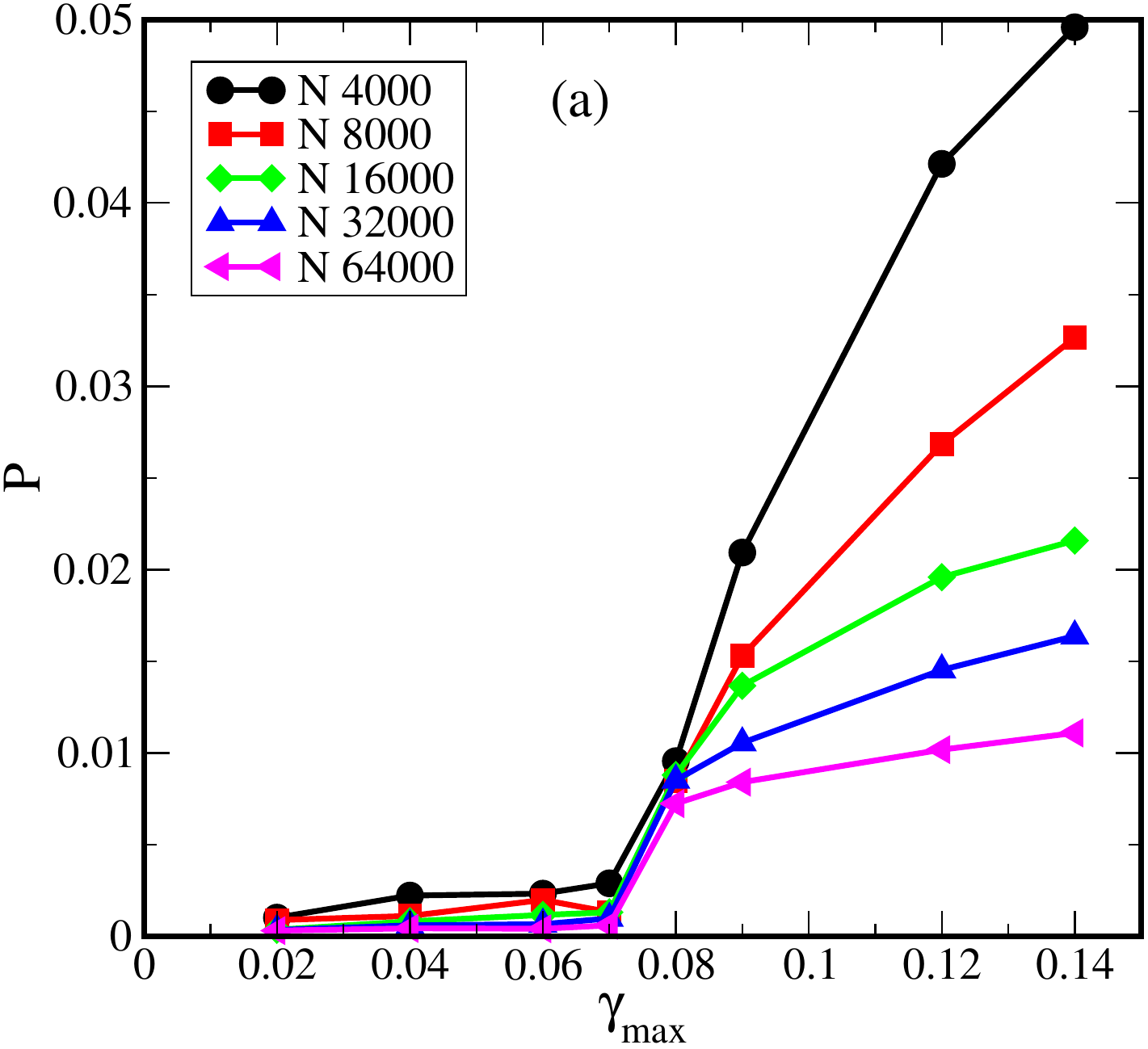}
\includegraphics[width=0.39\textwidth]{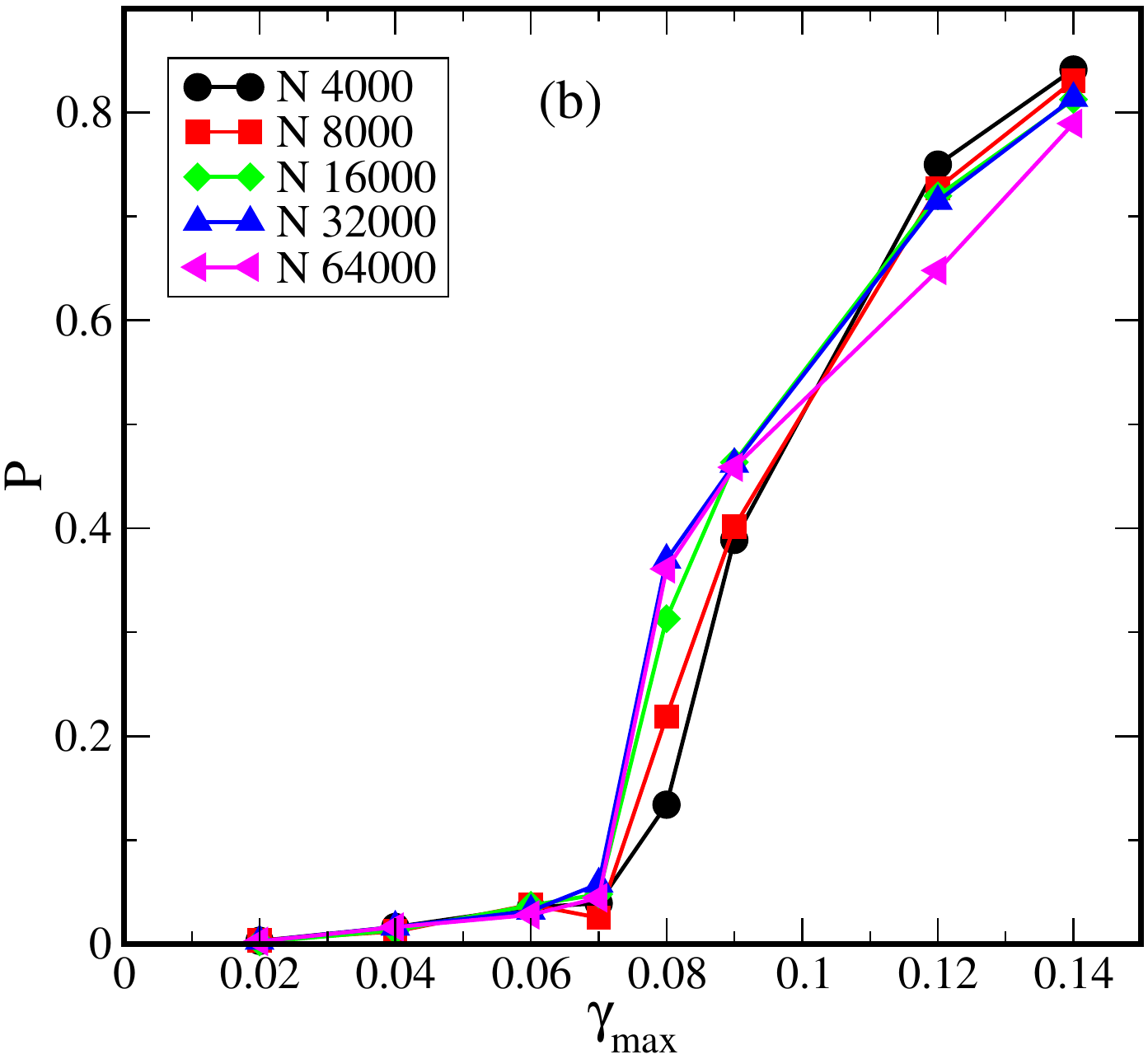}
\caption{Fraction of active particles $P$ {\it vs.} strain amplitude $\gamma_{max}$, for different system sizes, (a) for individual drop events, and (b) accumulated over the first quadrant of cycles of strain. }
\label{Fig:S:perco}
\end{figure}

We show in Fig. \ref{Fig:S:perco} the average fraction of active particles, or probability $P$, (ratio of the number of active particles and the total number of particles) for different system sizes, as a function of strain amplitude $\gamma_{max}$. The averages are performed over the first quadrant of the strain cycles. In  Fig. \ref{Fig:S:perco}(a) the probability $P$ is for individual drop events, averaged over all events, whereas in  Fig. \ref{Fig:S:perco} (b) $P$ is obtained for each cycle by accumulating all particles that are active in any of the drop events that occur, and the averaging is done over all cycles. In both cases, $P$ changes sharply across the yield strain, but with different system size dependence.

\end{document}